\documentclass[preprint,twocolumn]{aastex631}

\usepackage[utf8]{inputenc}
\usepackage{graphicx}
\usepackage{epsfig}
\usepackage{amsmath}
\usepackage{natbib}
\usepackage{footnote}
\usepackage{url}
\usepackage{appendix}

\graphicspath{{./}{figures/}}

\begin{document}
%\linenumbers

\title{IRAS\,00450+7401 and the mid-infrared fade/burst cycle of R Coronae Borealis-type stars}

\author[0000-0002-6023-7291]{William A. Burris}
\affiliation{Department of Astronomy, San Diego State University, 5500 Campanile Drive, San Diego, CA 92182, USA}

\author[0000-0001-9834-7579]{Carl Melis}
\affiliation{Center for Astrophysics and Space Sciences, University of California, San Diego, CA 92093-0424, USA}

\author[0000-0002-1276-1486]{Allen W.\ Shafter}
\affiliation{Department of Astronomy, San Diego State University, 5500 Campanile Drive, San Diego, CA 92182, USA}

\author[0000-0001-7482-5759]{Georgia V.\ Panopoulou}
\affiliation{Department of Space, Earth \& Environment, Chalmers University of Technology, SE-412 93 Gothenburg, Sweden}

\author[0000-0001-5058-1593]{E.\ L.\ Wright}
\affiliation{Physics and Astronomy Department, University of California, Los Angeles, CA 90095-1547, USA}

\author[0000-0003-0928-2000]{John Della Costa}
\affiliation{Department of Astronomy, San Diego State University, 5500 Campanile Drive, San Diego, CA 92182, USA}

\begin{abstract}
We present optical and infrared imaging and spectroscopy of the R Coronae Borealis-type
(R Cor Bor) star IRAS\,00450+7401. Optical spectra further confirm its classification as a cool R Cor Bor system,
having a hydrogen-deficient carbon star spectral sub-class of HdC5 or later. 
Mid-infrared spectroscopy reveals the typical $\sim$8\,$\mu$m ``hump'' seen in other R Cor Bor stars
and no other features.
A modern-epoch spectral energy distribution shows bright emission from hot dust having 
T$_{\rm dust}$$>$600\,K.
Historical infrared data reveal generally cooler dust color temperatures 
combined with long-term fading trends, but provide no discernible correlation between flux level and 
temperature. Investigating the most mid-infrared variable R Cor Bor stars found in 
{\it IRAS}, {\it AKARI}, and {\it WISE} data reveals similar fading trends, bursts that can 
show a factor of up to 10 change in flux density between epochs, and blackbody-fit 
dust color temperatures
that span 400-1300\,K. While some R Cor Bor stars such as IRAS\,00450+7401 appear to undergo fade/burst cycles in
the mid-infrared, significant gaps 
in temporal coverage prevent conclusively identifying any preferred timescale 
for their mid-infrared variability and circumstellar dust temperature changes. 

\end{abstract}


\section{Introduction}
R Coronae Borealis-type (R Cor Bor) stars are rare and luminous stellar systems that present
hydrogen-deficient and carbon-rich atmospheres, and experience irregular, sudden drops in
optical brightness by up to 8 or more magnitudes \citep{claytonCoronaeBorealisStars1996}.
The rarity of these systems (only $\sim$150 are currently known in the Galaxy;
\citealt{tisserandPlethoraNewCoronae2020,karambelkarCensusCoronaeBorealis2021})
suggest they may be the result of an atypical stellar evolutionary
pathway, or that they rapidly progress through this phase of evolution into more
typical stellar properties.

%%%%%%%%%%%%%%%%%%%%%%%%%%%%%%%%%%%%%%%%%%%%%%%%%%%%%%%%%%%%%%%%%%%%%%%%%%%%%%%%%%%
\begin{table*}[!ht]
\caption{{\large IRAS\,00450+7401 Epoch 2022 Photometry} \label{table:SED}}
\begin{center}
\begin{tabular}{ccccccc}
\hline 
\hline
Epoch & MJD & Bandpass & Magnitude & Flux (Jy) & Error (Jy) & Facility \\
\hline

2022-09-21 & 59843.36100 & 53.56\,$\mu$m & $-$ & $<$1.0 & $-$ & SOFIA \\
2022-09-15 & 59837.51315 & 37.12\,$\mu$m & $-$ & 0.63 & 0.15 & SOFIA \\
2022-05-17 & 59716.48368 & 34.68\,$\mu$m & $-$ & 0.64 & 0.07 & SOFIA \\
2022-09-15 & 59837.51013 & 33.43\,$\mu$m & $-$ & 0.65 & 0.13 & SOFIA \\
2022-05-17 & 59716.46903 & 33.57\,$\mu$m & $-$ & 0.57 & 0.19 & SOFIA \\
2022-05-17 & 59716.45410 & 31.36\,$\mu$m & $-$ & 0.74 & 0.09 & SOFIA \\
2022-05-17 & 59716.46903 & 25.24\,$\mu$m & $-$ & 0.93 & 0.11 & SOFIA \\
2022-05-17 & 59716.45410 & 19.7\,$\mu$m & $-$ & 1.64 & 0.11 & SOFIA \\
2022-02-19 & 59629.23297 & 11.09\,$\mu$m & $-$ & 3.39 & 0.25 & SOFIA \\
2022-02-17 & 59627.22448 & 6.35\,$\mu$m & $-$ & 4.06 & 0.22 & SOFIA \\
2022-01-27 & 59606.064 & 4.60\,$\mu$m & 4.64$\pm$0.06 & 2.38 & 0.13 & {\it WISE} \\
2022-01-27 & 59606.018 & 3.35\,$\mu$m & 5.90$\pm$0.05 & 1.34 & 0.06 & {\it WISE} \\
2022-02-15 & 59625.10686 & H & 9.2$\pm$0.1 & 0.22 & 0.02 & P200 \\
2022-02-15 & 59625.10972 & J & 10.40$\pm$0.03 & 0.112 & 0.003 & P200 \\
2022-03-02 & 59640.14699 & I & 12.25$\pm$0.04 & 33.1$\times$10$^{-3}$ & 1.2$\times$10$^{-3}$ & MLO \\
2022-03-02 & 59640.13435 & R & 13.56$\pm$0.03 & 11.0$\times$10$^{-3}$ & 0.3$\times$10$^{-3}$ & MLO \\
2022-03-02 & 59640.17218 & V & 15.16$\pm$0.03 & 3.27$\times$10$^{-3}$ & 0.01$\times$10$^{-3}$ & MLO \\
2022-03-02 & 59640.14699 & B & 17.04$\pm$0.08 & 0.63$\times$10$^{-3}$ & 0.04$\times$10$^{-3}$ & MLO \\
\hline
\end{tabular}
\\
{\it Note} $-$ {\it WISE} 3.35 and 4.60\,$\mu$m measurements are not color corrected $-$ corrections are on the order of 1\%. 
\end{center}
\end{table*}
%%%%%%%%%%%%%%%%%%%%%%%%%%%%%%%%%%%%%%%%%%%%%%%%%%%%%%%%%%%%%%%%%%%%%%%%%%%%%%%%%%%

There are two prevailing models for the origin of R Cor Bor stars. In what could be
considered the favored model, R Cor Bor stars are the remnants of a merger event
between two white dwarf stars where one object is CO-rich and the other He-rich
(e.g., \citealt{claytonWhatAreCoronae2012}).
Such a model would make R Cor Bor stars low-mass analogs of
supernova Type Ia progenitors \citep{fryerRoadUnderstandingType2008}
%(Fryer \& Diehl 2008) %Fryer \& Diehl 2008, ASPC 391, 335
where the components fail to
surpass the Chandrasekhar limit and explode.
This would be a rare and unusual stellar evolutionary pathway directly related
to binary star evolution.

Another model has R Cor Bor stars result from a final helium shell flash of a dying
giant star (e.g., \citealt{renziniEvolutionaryScenariosCrB1990}; \citealt{ibenOriginCrBOther1996}, and references therein). 
%Renzini 1990, ASPC 11, 549; Iben et al.\ 1996, ASPC 96, 409
This final flash
causes a significant portion of the stellar envelope to be lost and 
sufficient energy transfer to the remaining envelope and
core to have it expand into a cool supergiant-like object resembling R Cor Bor stars.
In this case the R Cor Bor phenomenon is a common event within the evolution of 
single low- to intermediate-mass stars and lasts only a short time.

The hallmark feature of R Cor Bor stars $-$ dramatic dimming events that can last
on the order of years $-$ is caused by clouds of circumstellar material
that cross our line of sight.
All R Cor Bor stars have an infrared excess caused by dust associated with the star, 
differentiating them from other hydrogen-deficient stars (e.g., 
\citealt{feastInfraredPhotometryCoronae1973,feast97,feastCoronaeBorealisStars1997}).
Understanding the nature and origin of the material that causes these dimming events can tell us 
about how these systems form and evolve.
%How this material forms, what it is made of, and how it evolves are areas of active research.
Evidence exists supporting dimming events being due to the ejection of clumps of 
atmospheric material (e.g., \citealt{claytonCircumstellarEnvironmentCoronae2011,claytonVariableWindsDust2013}).
Many R Cor Bor stars also have a 6--8\,$\mu$m ``hump'' in mid-infrared spectra 
thought to be due to carbon-dominated dust grains that could come from the carbon-rich
stellar atmosphere (\citealt{hechtDustCoronaeBorealis1984}; \citealt{wrightFractalDustGrains1989}; 
\citealt{lambertInfraredSpaceObservatory2001}; \citealt{garcia-hernandezDustCoronaeBorealis2013}).

%this paragraph gives background on the star
In this paper we discuss the candidate R Cor Bor system IRAS\,00450+7401 identified by 
\cite{tisserandTrackingCoronaeBorealis2012} and confirmed to have R Cor Bor star lightcurve behavior and near-infrared spectroscopic characteristics
by \cite{karambelkarCensusCoronaeBorealis2021}. IRAS\,00450+7401 came out of a $\sim$4.5 year 
dimming episode in early 2022 and optical spectra presented in this paper taken
near maximum light further confirm that it is indeed an R Cor Bor star. 

Section \ref{sec:2} discusses the suite of data we obtained for IRAS\,00450+7401 and Section 3 
reports results obtained from these data.
We present an optical/infrared spectral energy distribution (SED) for IRAS\,00450+7401 $-$ one of the most complete ever compiled in a year or less for an R Cor Bor star $-$ and characterize its infrared variability over an
$\approx$40~year time span using archival data. 
%We compare this infrared variability with R Cor Bor stars of similar variability and characterize it as one of the more variable in terms of its mid-infrared flux. %not really true
Section 4 synthesizes the new and 
historical data for IRAS\,00450+7401 and places them into context by comparison 
with other well-studied R Cor Bor stars.

%%%%%%%%%%%%%%%%%%%%%%%%%%%%%%%%%%%%%%%%%%%%%%%%%%%%%%%%%%%%%%%%%%%%%%%%%%%%%%%%%%%%

\begin{table*}[t!]
\caption{{\large IRAS\,00450+7401 MLO BVRI Photometry}
\label{table:vis_phot}}
\begin{center}
\begin{tabular}{cccccc}
\hline 
\hline
Epoch & MJD & Bandpass & Magnitude (Vega) & Flux (mJy) & Error (mJy)\\
\hline

2022-02-05 & 59615.16265 & I & 12.88$\pm$0.04 & 18.494 & 0.682 \\
2022-02-05 & 59615.16840 & R & 14.46$\pm$0.07 & 4.822 & 0.294 \\
2022-02-05 & 59615.16386 & V & 16.04$\pm$0.03 & 1.447 & 0.0465\\
2022-02-05 & 59615.15832 & B & 18.05$\pm$0.11 & 0.228 & 0.0258\\
2022-02-17 & 59627.13830 & I & 12.49$\pm$0.04 & 26.679 & 0.988\\
2022-02-17 & 59627.13524 & R & 13.81$\pm$0.03 & 8.812 & 0.246\\
2022-02-17 & 59627.17004 & V & 15.45$\pm$0.17 & 2.494 & 0.384 \\
2022-02-17 & $-$ & B & $-$ & $-$ & $-$ \\
2022-02-19 & 59629.09461 & I & 12.42$\pm$0.04 & 28.226 & 1.046 \\
2022-02-19 & 59629.08770 & R & 13.82$\pm$0.04 & 8.731 & 0.360 \\
2022-02-19 & 59629.11290 & V & 15.41$\pm$0.05 & 2.600 & 0.115 \\
2022-02-19 & 59629.10904 & B & 17.36$\pm$0.07 & 0.471 & 0.031  \\
2022-03-02 & 59640.14699 & I & 12.25$\pm$0.04 & 33.130 & 1.222\\
2022-03-02 & 59640.13435 & R & 13.56$\pm$0.03 & 11.040 & 0.306\\
2022-03-02 & 59640.17218 & V & 15.16$\pm$0.03 & 3.275 & 0.100\\
2022-03-02 & 59640.14699 & B & 17.04$\pm$0.08 & 0.633 & 0.044 \\
%2022-09-15 & $-$ & I & $-$ & $-$ & $-$ \\
2022-09-15 & 59845.39081 & R & 17.57$\pm$0.07 & 0.276 & 0.017 \\
2022-09-15 & 59845.38695 & V & 19.7$\pm$0.33 & 0.050 & 0.015 \\
2022-09-15 & $-$ & B & $-$ & $-$ & $-$ \\
\hline
\end{tabular}
\\
{\it Note} $-$ On 2022 February 17 UT and 2022 September 15 UT IRAS\,00450+7401 was not \\
detected in the \textit{B-}band due to bad weather and being too faint, respectively. 
%{\it Note} $-$ 
\end{center}
\end{table*}

%%%%%%%%%%%%%%%%%%%%%%%%%%%%%%%%%%%%%%%%%%%%%%%%%%%%%%%%%%%%%%%%%%%%%%%%%%%%%%%%%%%%%%%%%%%%%%%%%%%%

\section{Observations}
\label{sec:2}
%Discuss the observations and the methods of acquisition and reduction

We collected the following data for IRAS\,00450+7401: mid- to far-infrared photometry from 3--54\,$\mu$m, mid-infrared spectra from 5--14\,$\mu$m, near-infrared imaging in the \textit{J-} and \textit{H-}bands, optical spectra covering 3600--9000\,\AA , and optical photometric measurements in the \textit{B-}, \textit{V-}, \textit{R-}, and \textit{I-} bands.

\subsection{SOFIA Photometry}

Stratospheric Observatory for Infrared Astronomy (SOFIA) imaging observations were obtained for
IRAS\,00450+7401 on multiple flights in 2022 with the FORCAST \citep{herterDataReductionEarly2013} and HAWC+
\citep{harperHAWCFarInfraredCamera2018}
instruments. All observations at wavelengths longer than 11\,$\mu$m were performed
at an aircraft altitude of $\sim$43,000 feet.
Table \ref{table:SED} reports all observation dates and measured flux values.

While FORCAST hosts short- and long-wavelength channels (SW and LW, respectively) 
that are capable of simultaneously taking
images, most observations of IRAS\,00450+7401
utilized only a single channel at a time to avoid sensitivity loss from using
the dichroic. 
Dual-mode observations that simultaneously used both the SW and LW channels
were done for the 19.7 and 31.36\,$\mu$m pair and for the
25.24 and 33.57\,$\mu$m pair.
Observations were obtained in C2N (2-point, symmetrical)
chop–nod mode with a chop throw of $\pm$30$'$$'$ (resulting in one positive beam and
two negative beams with half the intensity of the positive beam). The telescope was
nodded in an ABBA pattern with nod offsets of 60$'$$'$. The chopping
and nodding position angles on the sky varied depending on the time of
observation, but always aligned such that chopping and nodding angles were matched.
Total on-source integration times ranged from roughly 1 to 8~minutes. 

Images are reduced and calibrated into physical flux units
by the SOFIA Data Cycle System (DCS) pipeline;
pipeline-reduced images have pixel scales of 0.768$'$$'$\,pixel$^{-1}$.
%pipeline version 2.5.0
These Level 3
data products are retrieved from the SOFIA archive service.
IRAS\,00450+7401 is detected as a point source in all FORCAST imaging
data with varying levels of signal-to-noise. As such, we attempted to
follow the guidance of the FORCAST Photometry data analysis 
cookbook\footnote{\url{https://github.com/SOFIAObservatory/Recipes/blob/master/FORCAST-photometry_detailed.ipynb}}
for measuring photometry. In short, this entails using a 12-pixel (9.22$'$$'$) radius aperture
(and larger sky annulus) to extract flux and estimate an uncertainty. 

As discussed in detail within the cookbook 
and in \citet{suInner25Au2017}, %Su, K. et al. 2017, AJ, 153, 226
using aperture radii other than 12 pixels is not recommended for extracting flux as the
point-spread function (PSF) is highly variable (even within a flight) and aperture
corrections cannot be reliably established.
IRAS\,00450+7401 is detected with high signal-to-noise ($\gtrsim$20 at the aperture,
meaning we take the signal in the aperture divided by the relative error in the 
photometry measurement as described
in the FORCAST Photometry cookbook)
for filters between 6--20\,$\mu$m, but is only detected at moderate or low signal-to-noise 
($<$15 at the aperture) for
filters longward of 20\,$\mu$m. Unfortunately, we found that standard 12-pixel radius
aperture extractions for these low- to moderate-signal-to-noise images resulted
in photometry dominated by noise; the resulting behavior of the source's flux between
30--40\,$\mu$m appeared quite erratic and unphysical.

The FORCAST Photometry cookbook suggests, following the work of \citet{suInner25Au2017}, that
PSF-fitting can be used to measure photometry when 
using high signal-to-noise observations of standard stars. 
Although the cookbook discusses this within the context of performing aperture corrections,
we decided to explore the possibility of using PSF-fitting to directly obtain photometric
measurements.
We use bright standard star images as an empirical model of the PSF, multiply them by a
constant factor to match the image of interest for IRAS\,00450+7401, and then perform the
regular cookbook aperture extraction on the scaled standard star image. Such aperture extractions
are not dominated by noise and should more faithfully measure the flux from the science target.

 All standard star observations obtained on the same night and with the same filter 
as observations for IRAS\,00450+7401 were downloaded. Standard star observations for the F315 filter
were not obtained on the same night as F315 observations for IRAS\,00450+7401, so we
used F315 standard star data from the subsequent night. All standard star images
are Level 3 data products and calibrated into physical flux units.

With in-house {\sf IDL} routines we modeled each image of IRAS\,00450+7401 with the same
filter image of the standard star effectively obtaining a scaling factor for
the standard star image. Aperture photometry with a 12-pixel radius aperture
was then performed on the scaled
standard star image while the uncertainty was calculated from the IRAS\,00450+7401 image
as specified in the cookbook. This approach yielded consistent results for filters
between 6--20\,$\mu$m and what appeared to be more accurate (at least more physically
reasonable) results for filters longward of 20\,$\mu$m.

For moderate- and especially low-signal-to-noise detections in FORCAST data
it seems standard aperture photometry with a 12-pixel radius aperture can produce
questionable results. We have found PSF-fitting using standard star observations
seems to produce more robust photometry values and strongly encourage such an
approach when analyzing FORCAST data.

Far-infrared imaging was obtained toward
IRAS\,00450+7401 with HAWC+.
Images were obtained at an aircraft altitude of 43,000~feet in total intensity
mode with the Band A filter and used Lissajous scan imaging. 
Data are reduced within DCS %can't find pipeline version..
and Level 3 calibrated
data products are obtained from the SOFIA science archive.
HAWC+ observations of IRAS\,00450+7401 were obtained in one of the final SOFIA flights.
During these last runs, HAWC+ experienced heightened consumption rates for its coolant
that sometimes impacted end-of-flight targets. Unfortunately, IRAS\,00450+7401 was one such
target and HAWC+ ran out of coolant after $\sim$5~minutes of on-source observation.
IRAS\,00450+7401 was not detected in this short HAWC+ sequence and we report an upper
limit in Table \ref{table:SED}.

\subsection{SOFIA Spectroscopy}

FORCAST spectroscopic observations were conducted with the G063 and G111
grisms in the SW channel using the 4.7$'$$'$ slit resulting in a 
spectral resolving power of $\sim$140.

G063 spectra were obtained on 2022 February 17 UT (MJD 59627.25208) %mid-time of 06:03
and covered roughly 5--8\,$\mu$m. During these observations
the aircraft started at an altitude of 37,000~feet and ended at 39,000~feet.
Chopping and nodding were performed in C2N mode. 
A total on-source integration time of 941~seconds was accrued with the G063 grism.
G111 spectra were obtained on 2022 February 19 UT (MJD 59629.24306) %mid-time of 05:50
and covered roughly 8.75--14\,$\mu$m. The aircraft altitude for these observations
was $\sim$39,000~feet and C2N chopping and nodding were performed.
G111 observations amassed a total on-source integration time of 469~seconds.

%pipeline version 2.3.0
Spectroscopic data are reduced, extracted, and calibrated within DCS
and Level 3 data products retrieved from the SOFIA science archive.
G063 data had manual telluric corrections performed with an ATRAN model
computed for a precipitable water vapor content of 3\,$\mu$m.
The final spectral signal-to-noise ratio is $\sim$60 at 7\,$\mu$m
and $\sim$20 at 11.5\,$\mu$m. However, this does not take into
account systematic noise from telluric correction (especially
apparent near the 9.5\,$\mu$m ozone feature).

%%%%%%%%%%%%%%%%%%%%%%%%%%%%%%%%%%%%%%%%%%%%%%%%%%%%%%%%%%%%%%%%%%%%%%%%%%%%%%%%%%%
\begin{figure}
 %\centering
 \hspace{-1cm}
 \includegraphics[width=95mm]{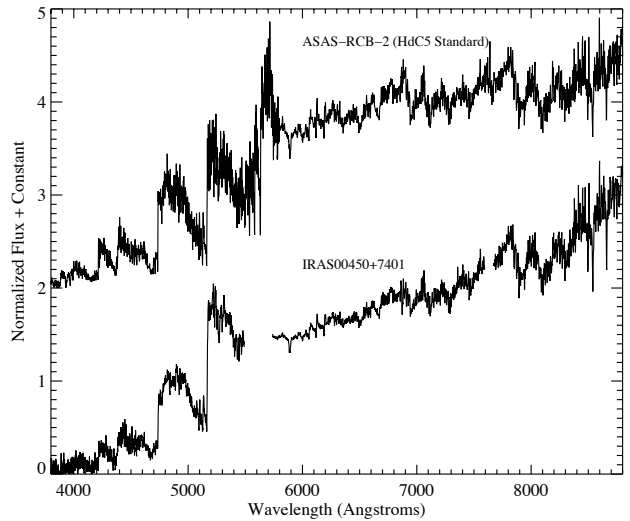}
 \caption{Kast optical spectra of IRAS\,00450+7401 taken when the
          source was recovering from a prolonged 
          deep dimming event and approximately 2.4~magnitudes below maximum
          light (see Section \ref{secphotstate}). Plotted above it is the earliest-type 
          spectral standard from \citet{crawfordSpectralClassificationSystem2022} that can match IRAS\,00450+7401
          (see Section \ref{secoptype}).
          Spectra are normalized at 4875\,\AA\ and the spectrum of ASAS-RCB-2 is vertically offset by
          a constant value for clarity. The spectrum of ASAS-2-RCB-2 is de-reddened and corrected for
          telluric absorption while the spectra
          of IRAS\,00450+7401 are not; the 7600\,\AA\ telluric feature 
          for IRAS\,00450+7401 is blanked out.
               }
\label{figkastspec}
\end{figure}

%%%%%%%%%%%%%%%%%%%%%%%%%%%%%%%%%%%%%%%%%%%%%%%%%%%%%%%%%%%%%%%%%%%%%%%%%%%%%%%%%%%

\subsection{{\it WISE} Photometry}

We incorporate NEOWISE 3.35 and 4.60\,$\mu$m data into our analysis (\citealt{wrightWidefieldInfraredSurvey2010}; \citealt{mainzerInitialPerformanceNEOWISE2014}). 

IRAS\,00450+7401 is saturated in NEOWISE single-frame images.
While the NEOWISE single-frame detection lists attempt to use
the PSF wings of saturated sources to recover valid
fluxes, the Explanatory Supplement\footnote{\url{https://wise2.ipac.caltech.edu/docs/release/neowise/expsup/sec2_1civa.html}} shows that there are still
biases for the brightest sources (like IRAS\,00450+7401). 

We perform an
additional correction to the NEOWISE-reported magnitudes retrieved for IRAS\,00450+7401
as follows:
W1/2$_{\mathrm{corr}}$ $=$ W1/2mpro + W1/2mcorr for W1/2mpro $<$ W1/2$_{\mathrm{satlimit}}$
and W1/2mcorr and W1/2$_{\mathrm{satlimit}}$ replaced by appropriate values for
either W1 or W2.
Uncertainties are the quadrature sum of the W1/2mpro catalog-quoted uncertainties
and the uncertainty on W1/2mcorr.

\subsection{Palomar Photometry}

Palomar Hale 200~inch (P200) near-infrared imaging was obtained for
IRAS\,00450+7401 on 2022 February 15 UT using the Wide-field Infrared
Camera (WIRC; \citealt{wilsonWideFieldInfraredCamera2003a}) and the 2048$\times$2048 Hawaii-II
detector that was installed in 2016.
%https://ui.adsabs.harvard.edu/abs/2003SPIE.4841..451W/abstract
Observations were obtained in the \textit{J-} and \textit{H-}bands, although only one
\textit{H-}band image was obtained as the target source was found to saturate
with the minimum exposure time of 0.92~seconds. Five un-dithered
exposures of 0.92~seconds each were obtained in the \textit{J-}band. Two of the \textit{J-}band
images have peak counts for IRAS\,00450+7401 that are well above the 1\% non-linearity
level (but still below full-well); we do not find results from these images to be
discrepant with those from the other images in the set and thus include them in our analysis
(excluding them does not have a significant impact on the final measured values).
%Linearity:  Detector is linear to 0.5% level up to 22,000 ADU, and 1% level up to 30,000 ADU. Full well is around 50,000 ADU.

WIRC images were reduced by subtracting a 0.92~second dark exposure and then
dividing by a twilight flat for each respective band.
Photometric extractions were performed for IRAS\,00450+7401 and several in-field reference stars
with 2MASS magnitudes brighter than $\sim$12 in the \textit{J-} and \textit{H-}bands.
An aperture that encloses $\sim$97\% of the flux was used for
all sources; resulting in $\lesssim$1\% variance within the set of five
\textit{J-}band images.

To obtain an \textit{H-}band magnitude for IRAS\,00450+7401, we performed PSF-fitting
with bright, unsaturated point sources serving as the PSF reference. PSF fits were
anchored on the unsaturated (and linear) wings of the source profile; saturated
and strongly non-linear pixels were masked. As a test case, we first attempted
a PSF fit for a field source also saturated in the \textit{H-}band image and found
excellent agreement with the 2MASS \textit{H-}band magnitude quoted for the source.
Despite this encouraging result, we adopt a generous uncertainty of 0.1~mag
for the IRAS\,00450+7401 \textit{H-}band measurement.

Final P200 infrared photometry measurements for IRAS\,00450+7401 are
reported in Table \ref{table:SED}.

%%%%%%%%%%%%%%%%%%%%%%%%%%%%%%%%%%%%%%%%%%%%%%%%%%%%%%%%%%%%%%%%%%%%%%%%%%%%%%%%%%%%

%             left panel - full SED+COMICS spectrum
%             right panel - COMICS spectrum only
\begin{figure}
 %\centering
 \hspace{-1cm}
 \includegraphics[width=97mm]{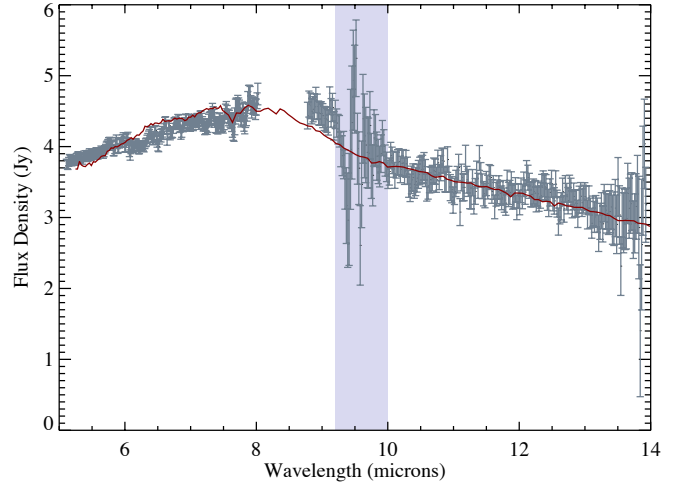}
 \caption{SOFIA FORCAST mid-infrared spectra of IRAS\,00450+7401
          plotted as grey error bars with a {\it Spitzer} IRS spectrum of the R Cor Bor star
          V739\,Sgr overplotted as the red  curve; other R Cor Bor spectra also provide a 
          similarly reasonable match, see discussion in Section \ref{sec:3.3}. The violet-shaded region indicates wavelengths affected by strong
          telluric absorption and where significant systematic errors are present in the data. Other telluric absorption is present throughout the data but has less of an overall impact.
          Spectra are not corrected
          for interstellar reddening or silicate absorption.  
               }
\label{figsofiaspec}
\end{figure}

%%%%%%%%%%%%%%%%%%%%%%%%%%%%%%%%%%%%%%%%%%%%%%%%%%%%%%%%%%%%%%%%%%%%%%%%%%%%%%%%%%%

%%%%%%%%%%%%%%%%%%%%%%%%%%%%%%%%%%%%%%%%%%%%%%%%%%%%%%%%%%%%%%%%%%%%%%%%%%%%%%%%%%%%
\begin{figure}
 \hspace{-0.3cm}
 \includegraphics[width=95mm]{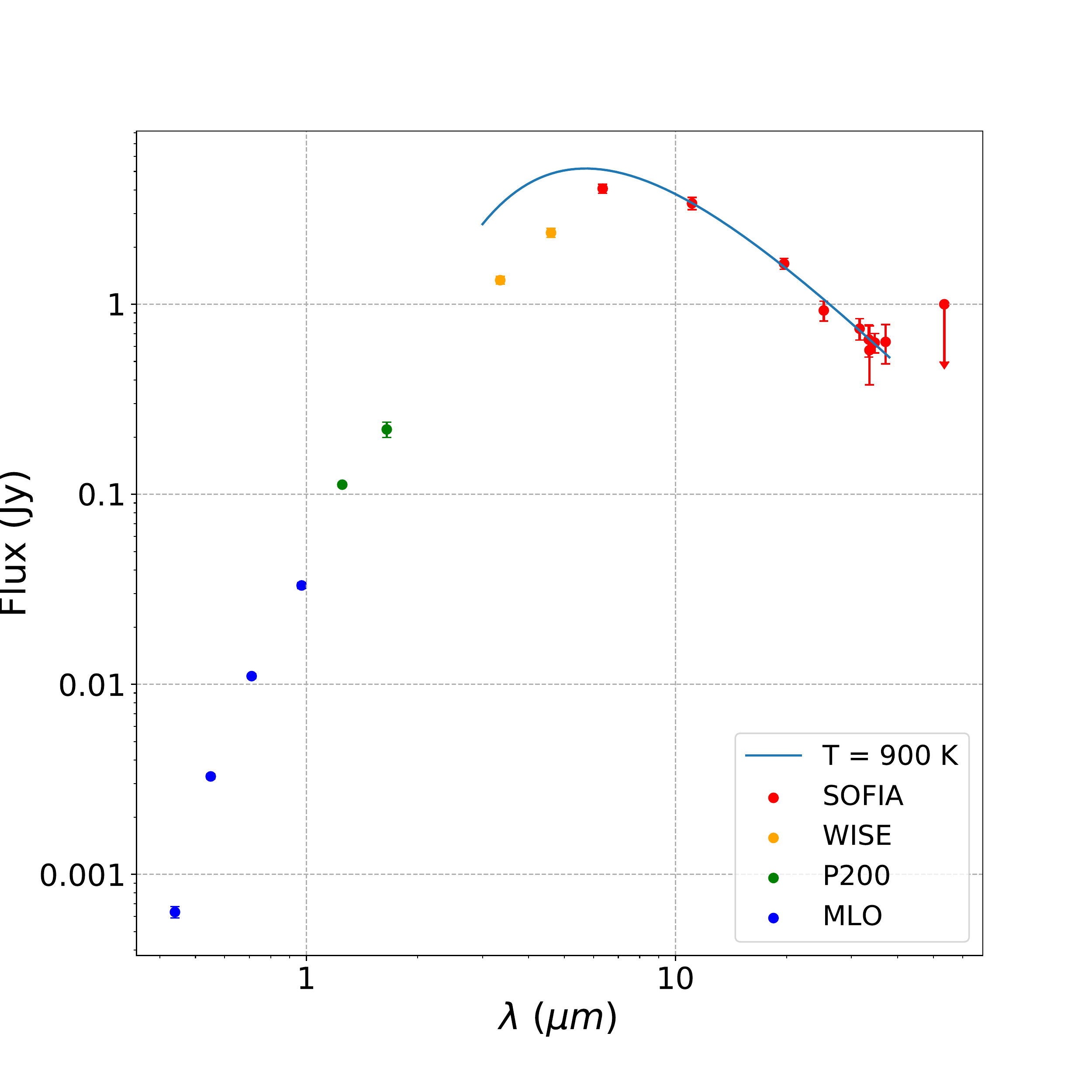}
 \caption{A 2022 epoch spectral energy distribution for IRAS\,00450+7401 covering 0.4--54\,$\mu$m (data point values are reported in Table \ref{table:SED}); {\it WISE} 3.35 and 4.60\,$\mu$m measurements are not color corrected as corrections are on the order of 1\%. The data point shown at 53.56\,$\mu$m is an upper limit. Overlaid is a blackbody computed with the best-fit temperature for the mid-infrared photometric data (data points $>$10\,$\mu$m and not including the 53.56\,$\mu$m upper limit); the fit allows for blackbody temperatures of 900$\pm$90\,K. While the blackbody fit is plotted down to $\approx$3\,$\mu$m, it is only informed
 by data with wavelengths $>$10\,$\mu$m. Section \ref{sec:3.3} discusses why we use such a fitting methodology even when it significantly over-predicts the flux from 3--6\,$\mu$m.
 }
 \label{fig:SED}
\end{figure}

%%%%%%%%%%%%%%%%%%%%%%%%%%%%%%%%%%%%%%%%%%%%%%%%%%%%%%%%%%%%%%%%%%%%%%%%%%%%%%%%%%%

\subsection{Optical Photometry}

Optical photometric measurements were taken with the San Diego State 40-inch Mount Laguna Observatory (MLO) telescope with the goal of characterizing the optical brightness of the star contemporaneous with the infrared measurements. %no known citation for the telescope/camera
\textit{B-}, \textit{V-}, \textit{R-}, and \textit{I-}band images were taken on four separate epochs in February--March of 2022, along with another epoch in September of 2022 (see Table \ref{table:vis_phot}). 

%%%%%%%%%%%%%%%%%%%%%%%%%%%%%%%%%%%%%%%%%%%%%%%%%%%%%%%%%%%%%%%%%%%%%%%%%%%%%%%%%%%%%%%%%%%%%%%%%%%%
\begin{table*}[!ht]
\caption{{\large IRAS\,00450+7401 Historical Infrared Photometry} \label{table:histSED}}
\begin{center}
\begin{tabular}{cccccc}
\hline 
\hline
Epoch & Bandpass & 	Magnitude (Vega) & Flux (Jy) & Error (Jy) & Facility \\
\hline
1983 & 60\,$\mu$m & $-$	& 0.47 & 0.07 & {\it IRAS} \\
1983 & 25\,$\mu$m & $-$ & 2.0 & 0.1 & {\it IRAS} \\
2010-02-12 & 22.09\,$\mu$m & 2.204$\pm$0.016 & 1.09 & 0.02 & {\it WISE} \\
2006 & 18\,$\mu$m & $-$ & 1.33 & 0.02 & {\it AKARI} \\
1983 & 12\,$\mu$m & $-$ & 3.80 & 0.15 & {\it IRAS} \\
2010-02-12 & 11.56\,$\mu$m & 3.069$\pm$0.010 & 1.72 & 0.02 & {\it WISE} \\
2006 & 9\,$\mu$m & $-$ & 2.38 & 0.02 & {\it AKARI} \\
2010-02-12 & 4.60\,$\mu$m & 4.834$\pm$0.061 & 2.0 & 0.1 & {\it WISE} \\
2010-02-12 & 3.35\,$\mu$m & 6.672$\pm$0.039 & 0.66 & 0.02 & {\it WISE} \\
1999-10-22 & K$_{\rm s}$ & 7.642$\pm$0.024 & 0.59 & 0.01 & 2MASS \\
1999-10-22 & H & 8.990$\pm$0.026 & 0.266 & 0.006 & 2MASS \\
1999-10-22 & J & 10.286$\pm$0.023 & 0.125 & 0.003 & 2MASS \\
\hline
\end{tabular}
\\
{\it Note} $-$ {\it IRAS}, {\it AKARI}, and {\it WISE} data are not color-corrected and are reported as retrieved from
their respective catalogs. Table \ref{table:color_correc} presents color-corrected space mission flux
values.
\end{center}
\end{table*}

%%%%%%%%%%%%%%%%%%%%%%%%%%%%%%%%%%%%%%%%%%%%%%%%%%%%%%%%%%%%%%%%%%%%%%%%%%%%%%%%%%%

Data were reduced within {\sf IRAF}, including performing bias subtraction and flat-fielding for each frame.
Differential photometry for each frame 
was obtained for the science target with the nearby star TYC 4308 226 1 serving 
as a reference source (adopted magnitudes for TYC 4308 226 1 can be found in Table \ref{table:ref_star}).
Typically three frames for each filter were obtained and the final magnitude
measurement is taken to be the average of the three frames with the uncertainty 
 given by the standard deviation.

Final magnitudes are converted into flux density values in Jy following
information provided in the \textit{NASA/IPAC Infrared Science Archive}.\footnote{\url{https://irsa.ipac.caltech.edu/data/SPITZER/docs/dataanalysistools/tools/pet/magtojy/ref.html}}.
The complete set of optical photometry can be found in Table \ref{table:vis_phot}.

\subsection{Optical Spectroscopy}
\label{seckast}

IRAS\,00450+7401 was observed through clear skies on 2022 January 28 UT at
Lick Observatory with the Kast Double Spectrograph mounted
on the Shane 3\,m telescope.
Before placing IRAS\,00450+7401 on the slit, we obtained guider camera images  of the science target and the nearby reference star TYC~4308 226 1 through
a Spinrad-R filter. From these
data we estimate an \textit{R-}band Vega magnitude for IRAS\,00450+7401 on 2022 January 28 UT (MJD 59607.12210)
of 14.4$\pm$0.1.

Kast observations simultaneously employed the blue and red arms
with light split by the d57 dichroic around 5700\,\AA.
After splitting, blue light was passed through the 600/4310 grism while
red light was passed through the 600/7500 grating. 
A slit width of 1.5$'$$'$ was used resulting in resolving powers of about
1100 in the blue spectra and 1750 in the red.

Kast data are reduced using standard
{\sf IRAF} long-slit tasks including bias subtraction, flat-fielding, wavelength calibration with
arc lamps, and instrumental response calibration via observations
of flux calibration standard stars. Arc lamp frames were not obtained close in time to science
frames and as such the zero-point of the wavelength scale is not accurate.
The final signal-to-noise ratio is $\sim$30 at 4900\,\AA\ in the blue spectrum
and $>$50 across the entire red spectrum.
We do not attempt to correct the Kast spectra for reddening.

%%%%%%%%%%%%%%%%%%%%%%%%%%%%%%%%%%%%%%%%%%%%%%%%%%%%%%%%%%%%%%%%%%%%%%%%%%%%%%%%%%%%
\begin{figure*}[!ht]
 \centering
 \includegraphics[width=180mm]{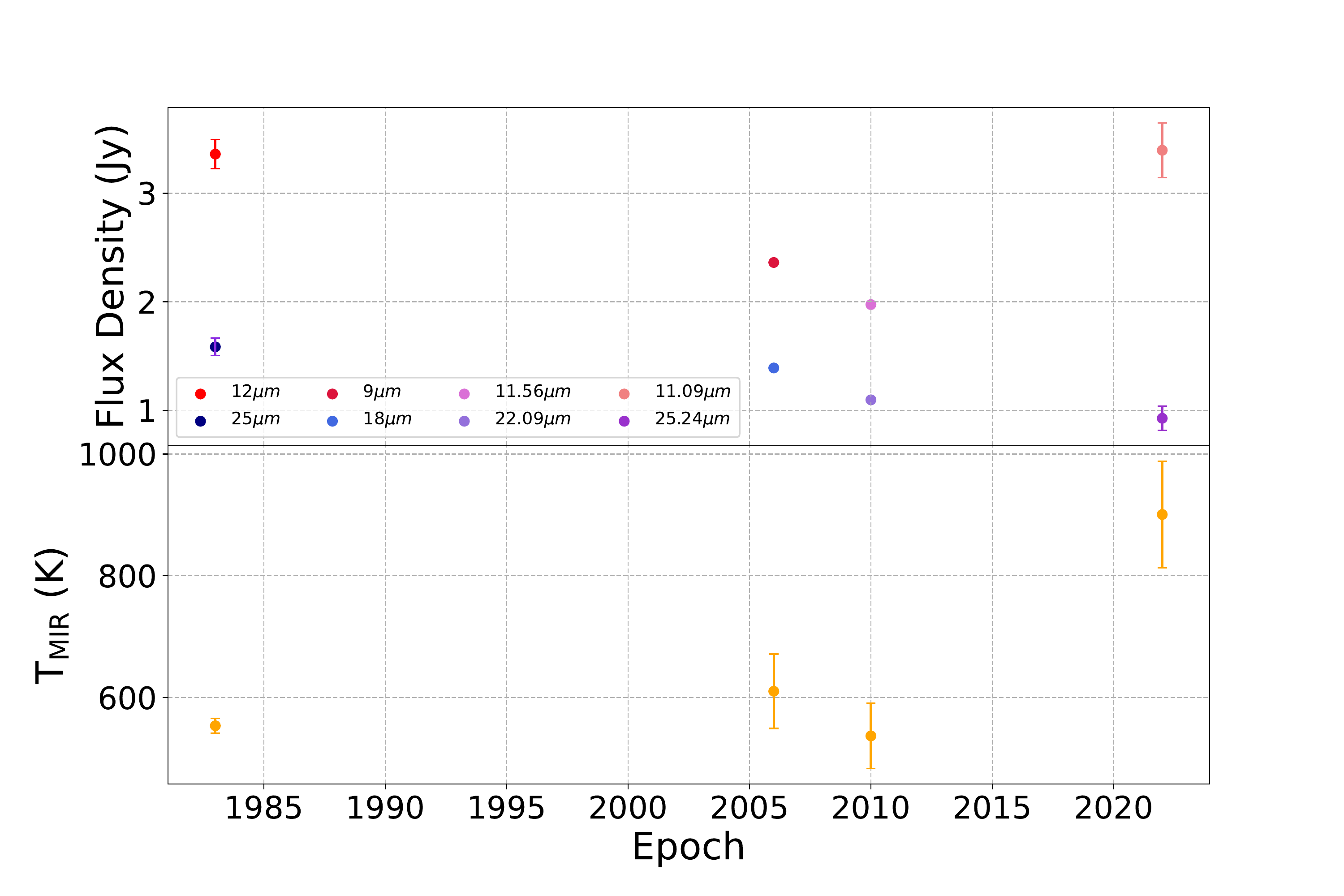}
 \caption{Top plot: Mid-infrared flux density of IRAS\,00450+7401 as a function of time for two different wavelengths per epoch. As measured by {\it AKARI}, {\it WISE}, {\it IRAS}, and {\it SOFIA} respectively, the red-hued (larger flux) points are at 9\,$\mu$m, 11.56\,$\mu$m, 12\,$\mu$m, and 11.09\,$\mu$m and the blue-hued (lower flux) points are at 18\,$\mu$m, 22.09\,$\mu$m, 25\,$\mu$m, and 25.24\,$\mu$m. 
 Historical photometric data points are color-corrected as described in Section \ref{sec:4}
 (see also Table \ref{table:color_correc}).
 Bottom plot: Blackbody-fit dust mid-infrared color temperature as a function of time (see Section \ref{sec:3.3})}.
 \label{fig:T_Over_t}
\end{figure*}

%%%%%%%%%%%%%%%%%%%%%%%%%%%%%%%%%%%%%%%%%%%%%%%%%%%%%%%%%%%%%%%%%%%%%%%%%%%%%%%%%%%

\section{Results}
\label{sec:3}

\subsection{Photometric State}
\label{secphotstate}

%When R Cor Bor stars go through optical dimming episodes their IR brightness remains relatively unchanged. 
We measured the brightness of IRAS\,00450+7401 in 
optical bands to assess what level of dimming the star was experiencing
as the infrared data were being collected. 
During the four initial epochs when visual data were taken, the star got consistently
brighter and marginally bluer. 
In nearly a month, from 2022 February 05 UT  to 2022 March 02 UT, the star brightened by $\sim$1 mag in the
\textit{B-}, 
\textit{V-}, 
and \textit{R-}bands and brightened by $\sim$0.6 mag in the \textit{I-}band (Table \ref{table:vis_phot}).
The R$-$I color of IRAS\,00450+7401 decreased from 1.56$\pm$0.08 to 1.31$\pm$0.05 over
the same time span. %this is barely significant, don't discuss further
%This is to be expected since the R Cor Bor dimming events are thought to be caused by a 
%build up of dust near the surface of the star, which affects shorter wavelength light greater. 
A fifth epoch of optical photometry on 2022 September 15 UT revealed that the star had
dipped again in brightness, fading by nearly 4 mag in the \textit{R-}band compared to the 2022 March epoch.

To better put MLO optical monitoring into context we utilized the \textit{dreams}\footnote{\url{https://dreams.anu.edu.au/monitoring/object?obj_name=WISE\%20J004822.34\%2B741757.4}}
monitoring webserver which contains brightness records for many R Cor Bor stars. 
ATLAS o-band monitoring \citep{tonryATLASHighcadenceAllsky2018} provides the most complete modern optical photometry data set for 
IRAS\,00450+7401 covering the times when the new data discussed above were being obtained and several
years leading up to their acquisition.
The ATLAS data set also cover what appears to be the brightest magnitudes and hence what
we tentatively interpret as maximum light for IRAS\,00450+7401. 
Around MJD 57966 (UT 2017 August 01),
IRAS\,00450+7401 had an o-band magnitude of 12.17$\pm$0.03. Following that maximum,
IRAS\,00450+7401 fell into a deep dimming state (o-band magnitudes consistently fainter than 15
and frequently $>$18) before beginning a resurgence around MJD 59490 (roughly 2021 October).
At the time of Kast observations (where we measured a rough \textit{R-}band magnitude of 14.4, see Section
\ref{seckast}) ATLAS measured an o-band magnitude of 14.70$\pm$0.03. The star continued to brighten
from that epoch consistent with MLO measurements, but unfortunately ATLAS did not visit the
star between 2022 March 02 UT and 2022 June 11 UT (presumably because it was at poor elevation). 
By the time ATLAS measurements for IRAS\,00450+7401 resumed, it had begun a new decline confirmed
with the faint state measured at MLO in September 2022.

%Use our data to estimate what state the star was in during infrared observations.
%https://dreams.anu.edu.au/monitoring/object?obj_name=WISE%20J004822.34%2B741757.4

\subsection{Spectroscopic Classification}
\label{secoptype}

In a recent work \citet{crawfordSpectralClassificationSystem2022} define
a spectral classification system for hydrogen-deficient carbon stars,
including R Cor Bor stars. While mostly based on observed spectral appearance,
the new classification system allows one to make a rough estimate
of the temperature for a star of known spectral type.

%HdC3 and HdC4 ruled out by mismatch in position of steep molecular shelves
Figure \ref{figkastspec} shows a comparison between optical spectra for
IRAS\,00450+7401 and the earliest-type matching standard from \citet{crawfordSpectralClassificationSystem2022}.
Earlier-type standards show a clear mis-match in the position of the steep molecular
shelves (e.g., near 4700\,\AA\ for IRAS\,00450+7401) or a lack of such features altogether, ruling out the HdC4 and earlier classifications,
(see Figure 4 of \citealt{crawfordSpectralClassificationSystem2022}). 
%Thus, we clearly identify IRAS\,00450+7401 as type HdC5 or later. 
%There is the possibility of IRAS\,00450+7401 belonging to a HC6 or HdC7 classification, but we cannot rule those out because of unknown veiling. 

%HdC5 is best match when de-reddening
Confounding a more exact spectral type estimate for IRAS\,00450+7401 is the fact that the
Kast spectra were obtained when the source was still recovering from a prolonged and deep
dimming event (see Section \ref{secphotstate}). The Kast spectra appear to have been obtained
when the source was $\sim$2.4~magnitudes below maximum light while all spectral standards
from \citet{crawfordSpectralClassificationSystem2022} were obtained near maximum light or at most 0.5~magnitudes below it.

Applying an A$_{\rm V}$$=$2.23~magnitude (assumed R$_{\rm V}$$=$3.1;
\citealt{hechtDustCoronaeBorealis1984} and references therein report R$_{\rm V}$ values
of $\sim$3 for R Cor Bor stars recovering from dimming events)
reddening correction to the Kast spectra for IRAS\,00450+7401 results in excellent agreement with
the HdC5 standard shown in Figure \ref{figkastspec}. We cannot verify such
a \textit{V-}band extinction value with the available photometric monitoring data for IRAS\,00450+7401.
The Shane-measured \textit{R-}band magnitude obtained on the same night as the Kast spectra
(Section \ref{seckast}) and
the ATLAS photometric monitoring around the same time (Section \ref{secphotstate})
suggest that A$_{\rm R}$$\sim$2.4
would be more appropriate (this would correspond to A$_{\rm V}$$=$3.06 for R$_{\rm V}$$=$3.1).
%the ATLAS $o$-band filter has a very wide bandpass that includes contributions from each of the V-, R-, and I-bands; it's effective wavelength is comparable to R-band

%but can't rule out HdC6 or HdC7 because of unknown veiling
R Cor Bor stars appear to have hot gas surrounding them that becomes
more apparent during declines (e.g., \citealt{claytonUnusualUltravioletChromospheric1992};  
Section 3.2.2 of \citealt{claytonCoronaeBorealisStars1996}; \citealt{kameswararaoRCoronaeBorealisStars2004}).
%https://ui.adsabs.harvard.edu/abs/1992ApJ...384L..19C/abstract
%https://ui.adsabs.harvard.edu/abs/2004MNRAS.355..855K/abstract
While reddening by dust can impact the overall slope of the Kast spectra, veiling
by such emission can act to weaken spectral features giving
an R Cor Bor
star the appearance of an earlier spectral type (e.g., \citealt{lambertHighResolutionSpectroscopy1990} and references therein).
%https://ui.adsabs.harvard.edu/abs/1990JApA...11..475L/abstract
What is not known is at what dimming state such veiling begins to become important
(for the case of IRAS\,00450+7401, such emission appears to be present
in near-infrared spectra taken during a dimming event observed by \citealt{karambelkarCensusCoronaeBorealis2021}).

Given the uncertainties around veiling for IRAS\,00450+7401 at the time of the Kast
spectroscopic observations, it is possible that it could be an HdC6 or even HdC7 type
star. The best way to resolve this issue would be to obtain a spectrum of
IRAS\,00450+7401 near maximum light, although it is not clear when that may happen again.
Until that time, we conservatively assign it a spectral type of HdC5 or later.

Based on Table 5 of \citet{crawfordSpectralClassificationSystem2022}, a spectral type of HdC5 or later
corresponds to a stellar effective temperature of $\lesssim$4500\,K. We thus
conclude that IRAS\,00450+7401 is a cool R Cor Bor star.

\vskip 0.3in
\subsection{Dust Characteristics}
\label{sec:3.3}

Mid-infrared spectroscopic data have been collected for $\sim$35 R Cor Bor stars with {\it ISO} and 
{\it Spitzer} (\citealt{lambertInfraredSpaceObservatory2001}; \citealt{kraemerCrBCandidatesSmall2005}; \citealt{garcia-hernandezAreC60Molecules2011}; \citealt{garcia-hernandezDustCoronaeBorealis2011}; \citealt{garcia-hernandezDustCoronaeBorealis2013}). These mid-infrared spectra
reveal a (likely carbon dust) ``hump'' feature between 6--8\,$\mu$m for most R Cor Bor stars, 
emission from polycyclic aromatic hydrocarbons for $\sim$10\% of R Cor Bor stars, and a smaller fraction still that may be host to $\mathrm{C_{60}}$ fullerenes.

The FORCAST spectra for IRAS\,00450+7401 demonstrate that this R Cor Bor star falls in line with the
majority of other R Cor Bor systems, displaying only a mild 6--8\,$\mu$m hump (Figure \ref{figsofiaspec}).
In an effort to better categorize IRAS\,00450+7401 amongst other R Cor Bor stars, we attempted to match
{\it Spitzer} IRS spectra presented by \citet{garcia-hernandezDustCoronaeBorealis2011}
with the FORCAST spectra for IRAS\,00450+7401. 
IRS spectra are obtained from the CASSIS database \citep{lebouteillerCASSISCornellAtlas2011} and are presented as
retrieved.
The best match overall is 
V739\,Sgr %HdC6: C25.5
which is shown in Figure \ref{figsofiaspec}. However, we note that 
V348\,Sgr, %no type
Z\,Umi, %HdC5: C25.5
RZ\,Nor, %HdC2 C22.5 Li
and WX\,Cra %HdC5: C25
also provide acceptable matches to the IRAS\,00450+7401 FORCAST spectra.
Table 4 of \citet{crawfordSpectralClassificationSystem2022} reports spectral types for 4 out of 5 of these
R Cor Bor stars, with all but RZ\,Nor probably belonging to cool subtypes (types of HdC5 or later).
This small sample possibly indicates that R Cor Bor dust does not depend on the temperature
class of the host star, a finding that should be more exhaustively investigated before being
considered conclusive.

Figure \ref{fig:SED} presents a 2022 epoch spectral energy distribution 
for IRAS\,00450+7401 with an accompanying blackbody fit for data points $>$10\,$\mu$m to 
roughly characterize the dust temperature as measured in the mid-infrared (T$_{\rm MIR}$). 
The data cover 0.44--53.56\,$\mu$m with 53.56\,$\mu$m being an upper limit and the blackbody fit yields a
temperature of T$_{\rm MIR}$$=$900$\pm$90\,K. 
When doing this blackbody fit we do not include data at wavelengths less than 10\,$\mu$m;
as seen in Figure \ref{fig:SED} the fit over-predicts the flux at 3--6\,$\mu$m wavelengths.
Fitting a 3-blackbody decomposition to all measurements from Table \ref{table:SED}
$-$ representing contributions from the star, an inner dust component, and an outer dust component $-$ results in temperatures of 2700, 1350, and 650\,K respectively. Such a fit reasonably reproduces
all Table \ref{table:SED} measurements and is not unique, but outer dust component temperatures 
$<$600\,K are ruled out.

While the 3-blackbody fitting procedure produces a better fit to the data, we will employ
throughout this paper the mid-infrared-only fitting approach. As will be seen in the next section,
this approach is necessary to maintain a consistent and homogeneous dust temperature measurement methodology when working with \textit{IRAS} and \textit{AKARI} data which only have data near 10 and 20\,$\mu$m. Temperatures estimated following the T$_{\rm MIR}$ approach are better viewed
as mid-infrared dust color temperatures since they miss contributions of dust components closer to the star.

Table 3 of \citet{garcia-hernandezDustCoronaeBorealis2011} presents multi-blackbody-fit dust temperatures
for $\sim$30 R Cor Bor stars observed with {\it Spitzer} IRS. They find inner dust components ranging in
temperature from 272-1600\,K and outer dust components (when present) ranging from 100-900\,K, thus placing the mid-infrared temperatures we report for IRAS\,00450+7401 somewhere near the high-end of
the mean.

\section{Discussion}
\label{sec:4}

%quick paragraph summarizing our basic optical/IR characterization of IRAS00450 in 2022 and how it looks pretty 'normal' for an R Cor Bor star.
Overall, we have found IRAS\,00450+7401 to share many of the same defining characteristics
of other R Cor Bor stars. Photometrically, it has spent the majority of the past $\sim$5~years in
a dim state, only resurfacing to near maximum light
for a brief time in the end of 2021/beginning of 2022 around when
observations for this paper were being collected. These characteristics mark IRAS\,00450+7401
as a relatively active R Cor Bor star. However, without the kind of 
longer historical monitoring that is available for 
other well-known R Cor Bor stars we cannot assess if this is a consistent feature for IRAS\,00450+7401.
Optical spectra taken near maximum light confirm it to be a cool R Cor Bor star with temperature
$<$5000\,K. Circumstellar dust around IRAS\,00450+7401 is of the expected carbon-rich composition
for an R Cor Bor system and presents in 2022 with an average-for-R Cor Bor stars temperature of $\gtrsim$600\,K.

%Then we move into its time-domain mid-IR properties and comparison to other R Cor Bors
Being relatively bright in the infrared, IRAS\,00450+7401 has detections from each of the 
{\it IRAS}, {\it AKARI}, and {\it WISE} all-sky
missions. We examined these data to see how the emission from dust 
around IRAS\,00450+7401 changes with time. Photometric measurements for each space mission need
to be color-corrected to arrive at accurate flux values, and these
color corrections are dependent on the intrinsic temperature of the source.
We fit blackbodies to the non-color corrected data as retrieved
from each mission's respective catalogs (Table \ref{table:histSED})
and used the resulting temperature to find the color correction factors as reported in online resources\footnote{\url{https://irsa.ipac.caltech.edu/IRASdocs/archives/colorcorr.html}} \footnote{\url{https://www.ir.isas.jaxa.jp/AKARI/Observation/support/IRC/IDUM/IRC_DUM.pdf}} \footnote{\url{https://wise2.ipac.caltech.edu/docs/release/allsky/expsup/sec4_4h.html\#FluxCC}};
color-corrected flux values are reported in Table \ref{table:color_correc}.

A blackbody is then fit to the color-corrected flux values to retrieve the mid-infrared dust 
color temperature (T$_{\rm MIR}$)
in each of the {\it IRAS}, {\it AKARI}, and {\it WISE} epochs.
{\it IRAS} and {\it AKARI} have only 2-3 data points available to inform blackbody
fits for each epoch and all with wavelengths $>$9\,$\mu$m. As such, we restrict our analysis
to wavelengths $>$9\,$\mu$m for all epochs as was done with the SOFIA data analysis and
fitting methodology (Figure \ref{fig:SED} and Section \ref{sec:3.3}).
With only 2-3 data points available per epoch, the blackbody
fit is over-constrained and the fitting routine cannot produce reliable error estimates
for the resulting color temperatures. We assign 10\% error bars to the three historical blackbody temperature fits in accordance with what was obtained for the more comprehensive
SOFIA data set.

Figure \ref{fig:T_Over_t} shows the mid-infrared flux and color temperatures 
for IRAS\,00450+7401 as a function of time.
Two curious features emerge from this figure. First, there does not seem to be an exact correlation
between the mid-infrared 
flux level and the blackbody-fit color temperature. While the SOFIA epoch is indeed the hottest
and brightest, the {\it IRAS} epoch is comparably bright but close to being the coolest.
Second, while sparsely populated, the overall trend of the mid-infrared light curve seems to suggest
a long-term fading of the circumstellar dust component followed by a return to a hot, bright phase.

%%%%%%%%%%%%%%%%%%%%%%%%%%%%%%%%%%%%%%%%%%%%%%%%%%%%%%%%%%%%%%%%%%%%%%%%%%%%%%%%%%%
\begin{figure*}[!ht]
 \centering
 \begin{minipage}[t!]{160mm}
  \includegraphics[width=150mm]{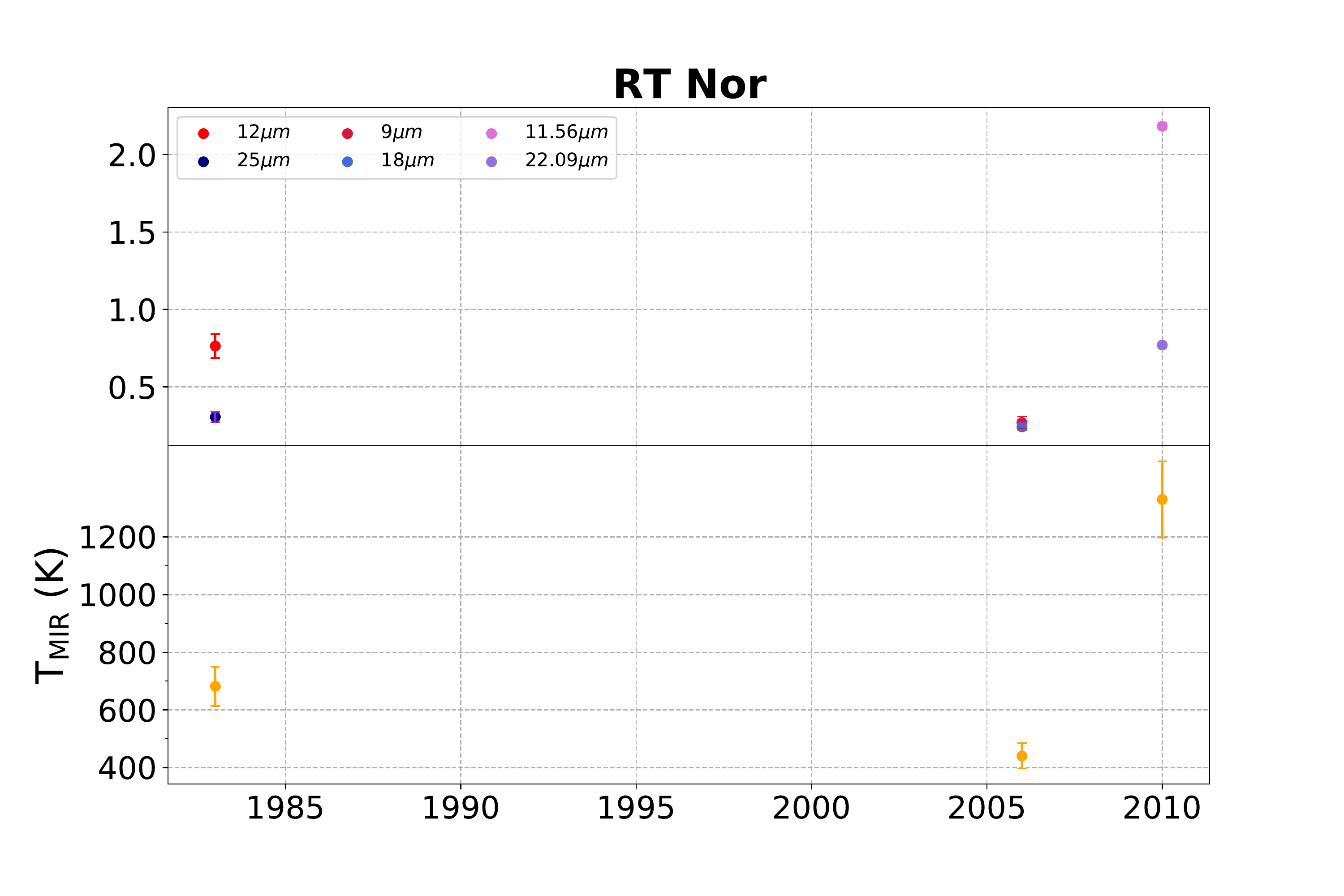}
 \end{minipage}
 \\*
 \begin{minipage}[b!]{160mm}
  \includegraphics[width=150mm]{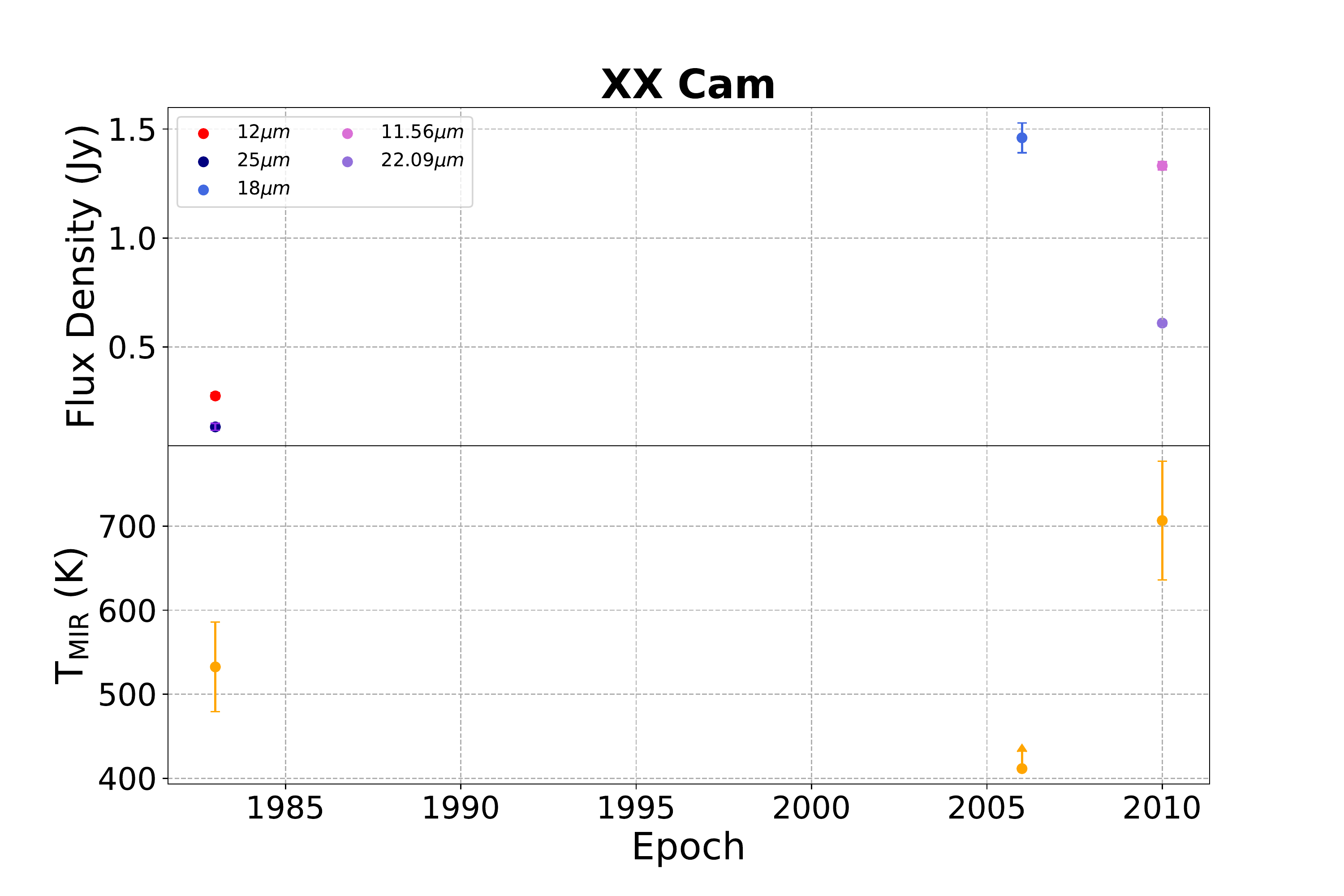}
 \end{minipage}
 \caption{ Mid-infrared flux and dust  color temperature variability for 
          RT Nor and XX Cam  (see Sections \ref{sec:3.3} and \ref{sec:4}); 
          symbols are as described in Figure
          \ref{fig:T_Over_t}. XX Cam only has an 18\,$\mu$m data point reported
          by {\it AKARI} and the dust color temperature is a limit assuming the 9\,$\mu$m flux is
          greater than or equal to the 18\,$\mu$m flux. Both stars exhibit marked dust brightening and
          heating events between 1983 and 2010.}
\label{fig:RT_XX_T_Over_t}
\end{figure*}

%%%%%%%%%%%%%%%%%%%%%%%%%%%%%%%%%%%%%%%%%%%%%%%%%%%%%%%%%%%%%%%%%%%%%%%%%%%%%%%%%%%

%%%%%%%%%%%%%%%%%%%%%%%%%%%%%%%%%%%%%%%%%%%%%%%%%%%%%%%%%%%%%%%%%%%%%%%%%%%%%%%%%%%
\begin{figure*}[!ht]
 \centering
 \begin{minipage}[t!]{160mm}
  \includegraphics[width=150mm]{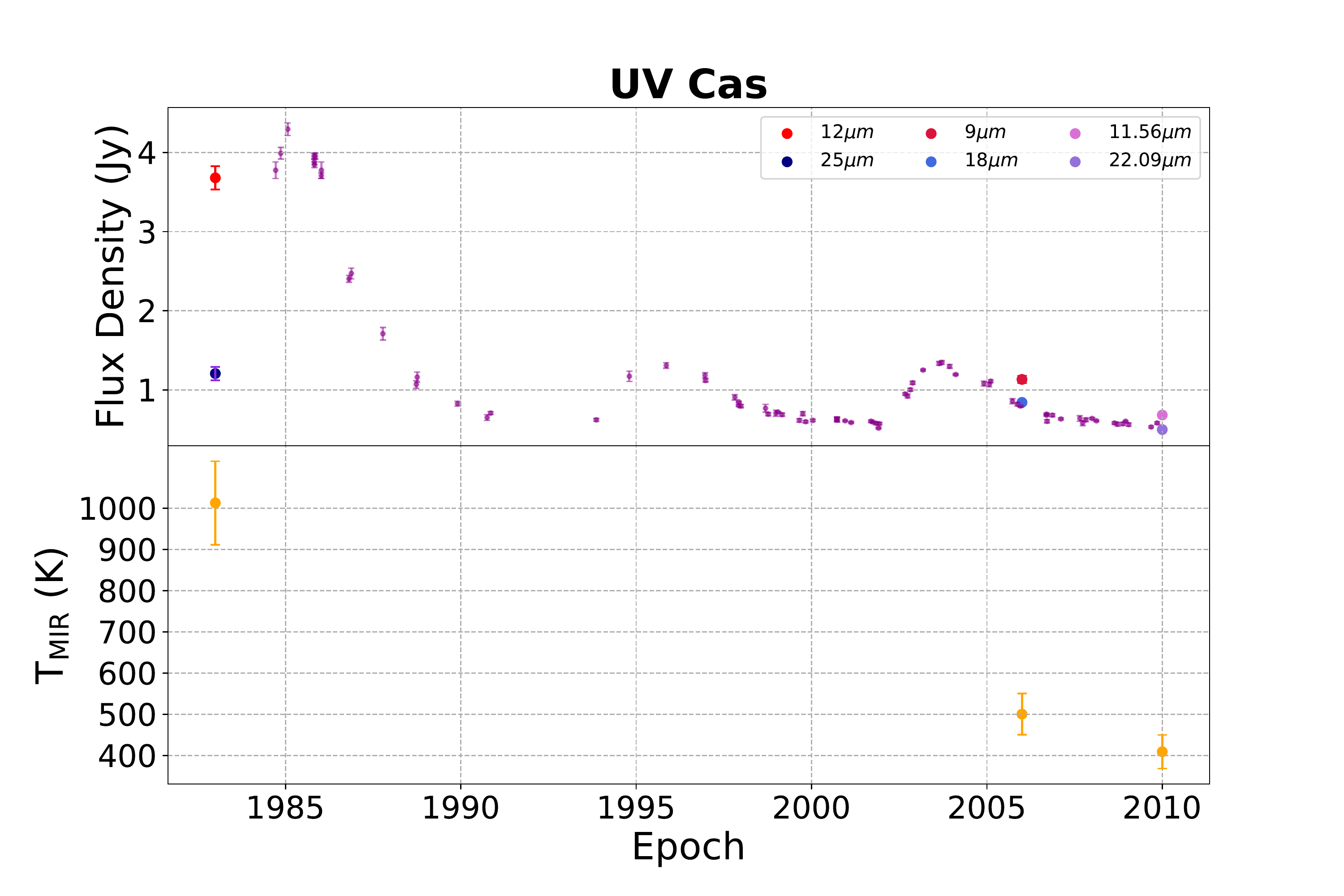}
 \end{minipage}
 \\*
 \begin{minipage}[b!]{160mm}
  \includegraphics[width=150mm]{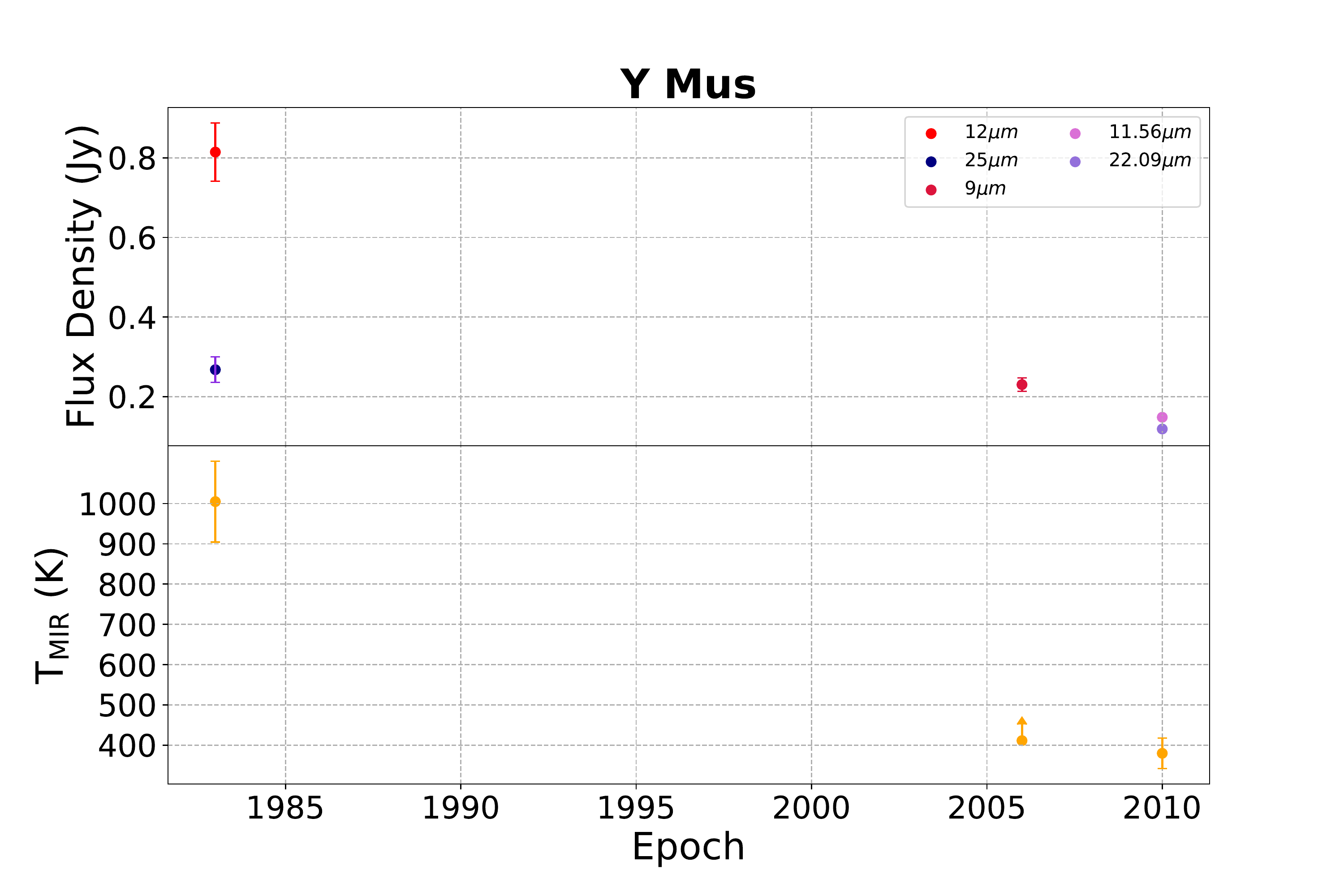}
 \end{minipage}
 \caption{Mid-infrared flux and dust color temperature variability for
          UV Cas and Y Mus (see Sections \ref{sec:3.3} and \ref{sec:4}}; 
          symbols are as described in Figure
          \ref{fig:T_Over_t}. Y Mus only has a 9\,$\mu$m data point reported
          by {\it AKARI} and the dust  color temperature is a limit assuming the 18\,$\mu$m flux is
          less than or equal to the 9\,$\mu$m flux. Raw L-band monitoring
          data for UV Cas from Table 1 of \citet{bogdanovIRPhotometryModels2010}
          are overplotted as small purple datapoints.
\label{fig:UV_T_Over_t}
\end{figure*}

%%%%%%%%%%%%%%%%%%%%%%%%%%%%%%%%%%%%%%%%%%%%%%%%%%%%%%%%%%%%%%%%%%%%%%%%%%%%%%%%%%%

%describe how we found RT Nor/etc by looking for >5x change in 12 micron flux between epochs
To put the mid-infrared variability of IRAS\,00450+7401 into context we sought out other
R Cor Bor stars that could be experiencing fading/brightening trends. We searched through
{\it IRAS}, {\it AKARI}, and {\it WISE} data for $\sim$120 Galactic R Cor Bor systems
and down-selected to the most dramatically variable sources which showed a factor of 5 or greater
change in flux between any two epochs. This approach yielded four R Cor Bor stars: RT Nor, XX Cam,
UV Cas, and Y Mus (all but XX Cam were similarly found to be highly variable in the
mid-infrared by \citealt{garcia-hernandezDustCoronaeBorealis2011}). 

Infrared data for each star are color corrected as described above
and blackbody fits are made to determine mid-infrared dust color temperatures (T$_{\rm MIR}$)
at different epochs. 
For Y Mus and XX Cam only one of the AKARI 9 or 18\,$\mu$m data points were available, 
but not both. Based on the observed $\sim$10 and $\sim$20\,$\mu$m data for other missions
and complete data sets for other stars, we assume as a limiting case that the 
9\,$\mu$m flux is greater than or equal to the 18\,$\mu$m flux and obtain a color
temperature lower limit.
Figures \ref{fig:RT_XX_T_Over_t} and \ref{fig:UV_T_Over_t} show the mid-infrared lightcurves
and estimated color temperatures for RT Nor, XX Cam, UV Cas, and Y Mus. Similar to the historical IRAS\,00450+7401 fits, we assign 10\% error bars to these blackbody-fit temperatures. Note that the {\it AKARI}
epoch falls close to when \citet{garcia-hernandezDustCoronaeBorealis2011} made {\it Spitzer}
observations of their R Cor Bor star sample which includes three of the four systems discussed in this
paragraph; 
{\it AKARI} data and associated blackbody fits are in good agreement with the results
of \citet{garcia-hernandezDustCoronaeBorealis2011}.

UV Cas and Y Mus show what appears to be prolonged fading trends, although significant gaps in temporal coverage could hide a variety of behaviors (see additional details on UV Cas below).
The color temperatures show a similar decay trend. This extends results by
\citet{garcia-hernandezDustCoronaeBorealis2011} for these two stars which were informed
by infrared data taken between 1983 and mid-2008.

RT Nor and XX Cam experience dramatic brightening episodes in the mid- to late-2000s, in the
case of RT Nor with a factor of $\sim$10 change in 10\,$\mu$m flux over only a few year time-span. 
These high-flux epochs are found to
be host to hot color temperatures, with values $>$1000\,K for RT Nor (we strongly suspect XX Cam
could have been similarly hot in the {\it AKARI} epoch, but the lack of a 9\,$\mu$m measurement prevents
confirming this suspicion).
\citet{garcia-hernandezDustCoronaeBorealis2011}
observed RT Nor with {\it Spitzer} in mid-2005 and did not observe XX Cam and as such missed
these burst-like events. 

Indeed, as Figure 12 of \citet{garcia-hernandezDustCoronaeBorealis2011}
shows, they did not identify any ``burst'' events between the {\it IRAS} and their
{\it Spitzer} epochs. It is probably the case that other R Cor Bor stars have experienced less dramatic
burst events than those seen for RT Nor and XX Cam (more akin to what
IRAS\,00450+7401 did between 2010 and 2022 as shown in Figure \ref{fig:T_Over_t});
investigating this will require a more exhaustive analysis of the available all-sky infrared databases.

%do all R Cor Bor stars have ~10 year fade/burst cycles?
Long term JHKL monitoring of 13 R Cor Bor stars
conducted by \citet{feast97} 
and \citet{bogdanovIRPhotometryModels2010} over the roughly 1980-2010 period
allows us to assess the above proposed mid-infrared
fade/burst hypothesis. L-band data probes dust emission (although there is some stellar contribution)
and is found to sometimes vary by up to 3~magnitudes over timescales of a few thousand days.
The only overlap between the five stars discussed thus far in this paper
and the JHKL monitoring sample is UV Cas. L-band photometry from \citet{bogdanovIRPhotometryModels2010} show
that UV Cas experiences burst episodes every $\sim$10~years (Figure \ref{fig:UV_T_Over_t}).
These bursts have variable maximum amplitudes and rise-to-maximum times that are on the order
of 2-3~years. Following
the bursts, the dust emission experiences an exponential-like decay over the next 6-8~years.
The times of maximum amplitude are not exactly 10~years apart with a peak in 1985 followed by
a peak in roughly 1996, and then another near the end of 2003.

While UV Cas presents a smoothly varying L-band lightcurve with readily identifiable fading/bursting
behavior,
many of the stars presented in \citet{feast97} show more stochastic
behavior and some even exhibit relatively flat L-band flux levels 
(the one star in \citealt{feast97} with probable periodicity in its L-band
dust emission level is R Coronae Borealis itself).
It is not clear what would cause some R Cor Bor stars and not others 
to have dust emission that goes through fade/burst
cycles; there does not appear to be any correlation with optical dimming events
(e.g., \citealt{feast97}; \citealt{garcia-hernandezDustCoronaeBorealis2011}).
Of course, for a sample of $\lesssim$10 stars as considered here and with only three epochs in the mid-infrared hosting a staggering coverage gap between 1983 and 2006, we 
should be careful not to hastily jump to any conclusions.

%seems like it is easier to see such cycles at 10 microns vs 3 microns
Comparing the L-band and 10--20\,$\mu$m lightcurves for UV Cas in Figure \ref{fig:UV_T_Over_t}
shows that the 20\,$\mu$m flux level seems to track
well with the L-band flux level. 
In 2006, during a time of decay after a burst episode, 
the 9\,$\mu$m flux is $\sim$50\% higher than the L-band measurement
while in 2010 the 11.56\,$\mu$m flux is comparable to the L-band flux level.
With the available data
it is not clear how the 10--20\,$\mu$m flux level would compare
to the L-band at burst peak amplitude (it is also not clear that the L-band
and mid-infrared wavelengths would necessarily peak at the same time).
Figure \ref{fig:SED} offers some insight, but only for the case of
IRAS\,00450+7401 which may or may not be applicable to other systems.

\iffalse
However, the 10\,$\mu$m flux level seems to be more pronounced;
e.g., in 1983 the 12\,$\mu$m flux is nearly the same as the maximum L-band flux measured in 1985 $-$
presumably the 12\,$\mu$m flux level in 1985 would have been higher than that seen in
the 1983 measurement (but this depends on the dust temperature in that epoch).
In 2006 the 9\,$\mu$m flux is $\sim$50\% higher than the L-band measurement. This implies
$-$ certainly for UV Cas and possibly more generally for R Cor Bor stars $-$ that dust emission variability
likely has higher amplitudes at wavelengths of 10\,$\mu$m and hence monitoring at 
such wavelengths may make it
easier to see putative fade/burst cycles that might be muted at shorter wavelengths.
\fi

Our modern and historical mid-infrared analysis of IRAS\,00450+7401 and a subset of other R Cor Bor 
stars make it clear that these systems would benefit from more frequent check-ups in the 
mid-infrared. % (certainly at 10\,$\mu$m). 
Consistent monitoring over a wide range of wavelengths
(e.g., 2--20\,$\mu$m) would elucidate if any fade/burst cycle is occurring for stars
that have not shown obvious evidence for such cycling in the L-band. 
Coverage over at least two wavelengths $-$
e.g., 11 and 18\,$\mu$m for ground-based observations through the N- and Q-band atmospheric windows $-$
would enable rough dust temperature measurements for each epoch along with monitoring flux levels
which would allow a new dimension of the dust characteristics of R Cor Bor stars to be probed.

%Bogdanov et al. (2010) survey of JHKLM for 25 years from 1984; only two stars in it, UV Cas and SU Tau
%also L-band photometric measurements conducted at the SAAO over several years (Feast et al. 1997) JHKL photometry is given for 11 R Coronae Borealis (RCB) stars
%  Feast+'97 Y Mus L-band 8.03 mag at epoch JD 2449886.3 / 19:12:0.00 UT June 17, 1995
%and of course the NEOWISE data which hasn't been investigated yet.. don't really want to go bothering with it for RT Nor et al.

%star temps
% UV Cas: 7200 K; HdC2
% RT Nor: 6700 K; HdC2
% XX Cam: ~6500 K; HdC2
% Y Mus : 7200 K; HdC0
%UV Cas and XX Cam are some of the least active R Cor Bors known in terms of dimming events
%   wonder what the mid-IR flux level of UV Cas is now..
%   did XX Cam have a dimming episode between 2000-2010?
%RT Nor was at maximum light when observed by Spitzer around AKARI epoch
%   did it have a dimming episode between 2005-2010?
%V348 Sgr (~20,000 K star) is one of the most active R Cor Bors known and has reasonably constant mid-IR flux in IRAS/AKARI/WISE

%\begin{acknowledgements}
\vskip 0.15in

We wish to thank Nicole Karnath and Bill Vacca for their excellent help in conducting SOFIA observations
and assessing the resultant data. 
We thank Courtney Crawford for providing HdC spectral standard data and general discussion about 
R Cor Bor star spectral types. We also thank the referee whose valuable insight has strengthened our work. 
This publication makes use of data products from the Near-Earth Object Wide-field Infrared Survey Explorer (NEOWISE), which is a joint project of the Jet Propulsion Laboratory/California Institute of Technology and the University of Arizona. NEOWISE is funded by the National Aeronautics and Space Administration. We acknowledge with thanks the variable star observations from the AAVSO International Database contributed by observers worldwide and used in this research. This work has made use of data from the European Space Agency (ESA) mission
{\it Gaia} (\url{https://www.cosmos.esa.int/gaia}), processed by the {\it Gaia}
Data Processing and Analysis Consortium (DPAC,
\url{https://www.cosmos.esa.int/web/gaia/dpac/consortium}). Funding for the DPAC
has been provided by national institutions, in particular the institutions
participating in the {\it Gaia} Multilateral Agreement.
Research at Lick Observatory is partially supported by a generous gift from Google.
This research has made use of NASA's Astrophysics Data System, the SIMBAD database,
and the VizieR service.

%\end{acknowledgements}

\facilities{SOFIA(FORCAST,HAWC+), Shane(Kast), SDSU 40-inch Mount Laguna Observatory telescope, {\it WISE}, {\it IRAS}, {\it AKARI}, Hale(WIRC)}

\bibliography{IRAS00450+740.bib}

\begin{thebibliography}{}
\expandafter\ifx\csname natexlab\endcsname\relax\def\natexlab#1{#1}\fi
\providecommand{\url}[1]{\href{#1}{#1}}
\providecommand{\dodoi}[1]{doi:~\href{http://doi.org/#1}{\nolinkurl{#1}}}
\providecommand{\doeprint}[1]{\href{http://ascl.net/#1}{\nolinkurl{http://ascl.net/#1}}}
\providecommand{\doarXiv}[1]{\href{https://arxiv.org/abs/#1}{\nolinkurl{https://arxiv.org/abs/#1}}}

\bibitem[{Bogdanov {et~al.}(2010)Bogdanov, Taranova, \&
  Shenavrin}]{bogdanovIRPhotometryModels2010}
Bogdanov, M.~B., Taranova, O.~G., \& Shenavrin, V.~I. 2010, Astronomy Reports,
  54, 620, \dodoi{10.1134/S1063772910070061}

\bibitem[{Clayton(1996)}]{claytonCoronaeBorealisStars1996}
Clayton, G.~C. 1996, Publications of the Astronomical Society of the Pacific,
  108, 225, \dodoi{10.1086/133715}

\bibitem[{Clayton(2012)}]{claytonWhatAreCoronae2012}
---. 2012, What are the {R} {Coronae} {Borealis} {Stars}?,  arXiv.
\newblock \url{http://arxiv.org/abs/1206.3448}

\bibitem[{Clayton {et~al.}(2013)Clayton, Geballe, \&
  Zhang}]{claytonVariableWindsDust2013}
Clayton, G.~C., Geballe, T.~R., \& Zhang, W. 2013, The Astronomical Journal,
  146, 23, \dodoi{10.1088/0004-6256/146/2/23}

\bibitem[{Clayton {et~al.}(1992)Clayton, Whitney, Stanford, Drilling, \&
  Judge}]{claytonUnusualUltravioletChromospheric1992}
Clayton, G.~C., Whitney, B.~A., Stanford, S.~A., Drilling, J.~S., \& Judge,
  P.~G. 1992, The Astrophysical Journal, 384, L19, \dodoi{10.1086/186253}

\bibitem[{Clayton {et~al.}(2011)Clayton, Sugerman, Stanford, Whitney, Honor,
  Babler, Barlow, Gordon, Andrews, Geballe, Bond, De~Marco, Lawson, Sibthorpe,
  Olofsson, Polehampton, Gomez, Matsuura, Hargrave, Ivison, Wesson, Leeks,
  Swinyard, \& Lim}]{claytonCircumstellarEnvironmentCoronae2011}
Clayton, G.~C., Sugerman, B. E.~K., Stanford, S.~A., {et~al.} 2011, The
  Astrophysical Journal, 743, 44, \dodoi{10.1088/0004-637X/743/1/44}

\bibitem[{Collaboration {et~al.}(2016)Collaboration, Prusti, de~Bruijne, Brown,
  Vallenari, Babusiaux, Bailer-Jones, Bastian, Biermann, Evans, Eyer, Jansen,
  Jordi, Klioner, Lammers, Lindegren, Luri, Mignard, Milligan, Panem,
  Poinsignon, Pourbaix, Randich, Sarri, Sartoretti, Siddiqui, Soubiran,
  Valette, van Leeuwen, Walton, Aerts, Arenou, Cropper, Drimmel, Høg, Katz,
  Lattanzi, O'Mullane, Grebel, Holland, Huc, Passot, Bramante, Cacciari,
  Castañeda, Chaoul, Cheek, De~Angeli, Fabricius, Guerra, Hernández,
  Jean-Antoine-Piccolo, Masana, Messineo, Mowlavi, Nienartowicz,
  Ordóñez-Blanco, Panuzzo, Portell, Richards, Riello, Seabroke, Tanga,
  Thévenin, Torra, Els, Gracia-Abril, Comoretto, Garcia-Reinaldos, Lock,
  Mercier, Altmann, Andrae, Astraatmadja, Bellas-Velidis, Benson, Berthier,
  Blomme, Busso, Carry, Cellino, Clementini, Cowell, Creevey, Cuypers,
  Davidson, De~Ridder, de~Torres, Delchambre, Dell'Oro, Ducourant, Frémat,
  García-Torres, Gosset, Halbwachs, Hambly, Harrison, Hauser, Hestroffer,
  Hodgkin, Huckle, Hutton, Jasniewicz, Jordan, Kontizas, Korn, Lanzafame,
  Manteiga, Moitinho, Muinonen, Osinde, Pancino, Pauwels, Petit, Recio-Blanco,
  Robin, Sarro, Siopis, Smith, Smith, Sozzetti, Thuillot, van Reeven, Viala,
  Abbas, Abreu~Aramburu, Accart, Aguado, Allan, Allasia, Altavilla, Álvarez,
  Alves, Anderson, Andrei, Anglada~Varela, Antiche, Antoja, Antón, Arcay,
  Atzei, Ayache, Bach, Baker, Balaguer-Núñez, Barache, Barata, Barbier,
  Barblan, Baroni, Barrado~y Navascués, Barros, Barstow, Becciani, Bellazzini,
  Bellei, Bello~García, Belokurov, Bendjoya, Berihuete, Bianchi, Bienaymé,
  Billebaud, Blagorodnova, Blanco-Cuaresma, Boch, Bombrun, Borrachero,
  Bouquillon, Bourda, Bouy, Bragaglia, Breddels, Brouillet, Brüsemeister,
  Bucciarelli, Budnik, Burgess, Burgon, Burlacu, Busonero, Buzzi, Caffau,
  Cambras, Campbell, Cancelliere, Cantat-Gaudin, Carlucci, Carrasco,
  Castellani, Charlot, Charnas, Charvet, Chassat, Chiavassa, Clotet, Cocozza,
  Collins, Collins, Costigan, Crifo, Cross, Crosta, Crowley, Dafonte, Damerdji,
  Dapergolas, David, David, De~Cat, de~Felice, de~Laverny, De~Luise, De~March,
  de~Martino, de~Souza, Debosscher, del Pozo, Delbo, Delgado, Delgado,
  di~Marco, Di~Matteo, Diakite, Distefano, Dolding, Dos~Anjos, Drazinos,
  Durán, Dzigan, Ecale, Edvardsson, Enke, Erdmann, Escolar, Espina, Evans,
  Eynard~Bontemps, Fabre, Fabrizio, Faigler, Falcão, Farràs~Casas, Faye,
  Federici, Fedorets, Fernández-Hernández, Fernique, Fienga, Figueras,
  Filippi, Findeisen, Fonti, Fouesneau, Fraile, Fraser, Fuchs, Furnell, Gai,
  Galleti, Galluccio, Garabato, García-Sedano, Garé, Garofalo, Garralda,
  Gavras, Gerssen, Geyer, Gilmore, Girona, Giuffrida, Gomes, González-Marcos,
  González-Núñez, González-Vidal, Granvik, Guerrier, Guillout, Guiraud,
  Gúrpide, Gutiérrez-Sánchez, Guy, Haigron, Hatzidimitriou, Haywood, Heiter,
  Helmi, Hobbs, Hofmann, Holl, Holland, Hunt, Hypki, Icardi, Irwin, Jevardat~de
  Fombelle, Jofré, Jonker, Jorissen, Julbe, Karampelas, Kochoska, Kohley,
  Kolenberg, Kontizas, Koposov, Kordopatis, Koubsky, Kowalczyk, Krone-Martins,
  Kudryashova, Kull, Bachchan, Lacoste-Seris, Lanza, Lavigne,
  Le~Poncin-Lafitte, Lebreton, Lebzelter, Leccia, Leclerc, Lecoeur-Taibi,
  Lemaitre, Lenhardt, Leroux, Liao, Licata, Lindstrøm, Lister, Livanou, Lobel,
  Löffler, López, Lopez-Lozano, Lorenz, Loureiro, MacDonald,
  Magalhães~Fernandes, Managau, Mann, Mantelet, Marchal, Marchant, Marconi,
  Marie, Marinoni, Marrese, Marschalkó, Marshall, Martín-Fleitas, Martino,
  Mary, Matijevič, Mazeh, McMillan, Messina, Mestre, Michalik, Millar,
  Miranda, Molina, Molinaro, Molinaro, Molnár, Moniez, Montegriffo, Monteiro,
  Mor, Mora, Morbidelli, Morel, Morgenthaler, Morley, Morris, Mulone, Muraveva,
  Musella, Narbonne, Nelemans, Nicastro, Noval, Ordénovic, Ordieres-Meré,
  Osborne, Pagani, Pagano, Pailler, Palacin, Palaversa, Parsons, Paulsen,
  Pecoraro, Pedrosa, Pentikäinen, Pereira, Pichon, Piersimoni, Pineau, Plachy,
  Plum, Poujoulet, Prša, Pulone, Ragaini, Rago, Rambaux, Ramos-Lerate,
  Ranalli, Rauw, Read, Regibo, Renk, Reylé, Ribeiro, Rimoldini, Ripepi, Riva,
  Rixon, Roelens, Romero-Gómez, Rowell, Royer, Rudolph, Ruiz-Dern, Sadowski,
  Sagristà~Sellés, Sahlmann, Salgado, Salguero, Sarasso, Savietto, Schnorhk,
  Schultheis, Sciacca, Segol, Segovia, Segransan, Serpell, Shih, Smareglia,
  Smart, Smith, Solano, Solitro, Sordo, Soria~Nieto, Souchay, Spagna, Spoto,
  Stampa, Steele, Steidelmüller, Stephenson, Stoev, Suess, Süveges, Surdej,
  Szabados, Szegedi-Elek, Tapiador, Taris, Tauran, Taylor, Teixeira, Terrett,
  Tingley, Trager, Turon, Ulla, Utrilla, Valentini, van Elteren, Van~Hemelryck,
  van Leeuwen, Varadi, Vecchiato, Veljanoski, Via, Vicente, Vogt, Voss,
  Votruba, Voutsinas, Walmsley, Weiler, Weingrill, Werner, Wevers, Whitehead,
  Wyrzykowski, Yoldas, Žerjal, Zucker, Zurbach, Zwitter, Alecu, Allen,
  Allende~Prieto, Amorim, Anglada-Escudé, Arsenijevic, Azaz, Balm, Beck,
  Bernstein, Bigot, Bijaoui, Blasco, Bonfigli, Bono, Boudreault, Bressan,
  Brown, Brunet, Bunclark, Buonanno, Butkevich, Carret, Carrion, Chemin,
  Chéreau, Corcione, Darmigny, de~Boer, de~Teodoro, de~Zeeuw, Delle~Luche,
  Domingues, Dubath, Fodor, Frézouls, Fries, Fustes, Fyfe, Gallardo, Gallegos,
  Gardiol, Gebran, Gomboc, Gómez, Grux, Gueguen, Heyrovsky, Hoar, Iannicola,
  Isasi~Parache, Janotto, Joliet, Jonckheere, Keil, Kim, Klagyivik, Klar,
  Knude, Kochukhov, Kolka, Kos, Kutka, Lainey, LeBouquin, Liu, Loreggia,
  Makarov, Marseille, Martayan, Martinez-Rubi, Massart, Meynadier, Mignot,
  Munari, Nguyen, Nordlander, Ocvirk, O'Flaherty, Olias~Sanz, Ortiz, Osorio,
  Oszkiewicz, Ouzounis, Palmer, Park, Pasquato, Peltzer, Peralta, Péturaud,
  Pieniluoma, Pigozzi, Poels, Prat, Prod'homme, Raison, Rebordao, Risquez,
  Rocca-Volmerange, Rosen, Ruiz-Fuertes, Russo, Sembay, Serraller~Vizcaino,
  Short, Siebert, Silva, Sinachopoulos, Slezak, Soffel, Sosnowska, Straižys,
  ter Linden, Terrell, Theil, Tiede, Troisi, Tsalmantza, Tur, Vaccari, Vachier,
  Valles, Van~Hamme, Veltz, Virtanen, Wallut, Wichmann, Wilkinson, Ziaeepour,
  \& Zschocke}]{collaborationGaiaMission2016}
Collaboration, G., Prusti, T., de~Bruijne, J. H.~J., {et~al.} 2016, Astronomy
  and Astrophysics, 595, A1, \dodoi{10.1051/0004-6361/201629272}

\bibitem[{Crawford {et~al.}(2022)Crawford, Tisserand, Clayton, Soon, Bessell,
  Wood, \& Garcia-Hernandez}]{crawfordSpectralClassificationSystem2022}
Crawford, C.~L., Tisserand, P., Clayton, G.~C., {et~al.} 2022, A {Spectral}
  {Classification} {System} for {Hydrogen}-deficient {Carbon} {Stars}, Tech.
  rep.
\newblock \url{https://ui.adsabs.harvard.edu/abs/2022arXiv221004416C}

\bibitem[{Feast(1997)}]{feastCoronaeBorealisStars1997}
Feast, M.~W. 1997, Monthly Notices of the Royal Astronomical Society, 285, 339,
  \dodoi{10.1093/mnras/285.2.339}

\bibitem[{{Feast} {et~al.}(1997){Feast}, {Carter}, {Roberts}, {Marang}, \&
  {Catchpole}}]{feast97}
{Feast}, M.~W., {Carter}, B.~S., {Roberts}, G., {Marang}, F., \& {Catchpole},
  R.~M. 1997, \mnras, 285, 317, \dodoi{10.1093/mnras/285.2.317}

\bibitem[{Feast \& Glass(1973)}]{feastInfraredPhotometryCoronae1973}
Feast, M.~W., \& Glass, I.~S. 1973, Monthly Notices of the Royal Astronomical
  Society, 161, 293, \dodoi{10.1093/mnras/161.3.293}

\bibitem[{Fryer \& Diehl(2008)}]{fryerRoadUnderstandingType2008}
Fryer, C.~L., \& Diehl, S. 2008, 391, 335.
\newblock \url{https://ui.adsabs.harvard.edu/abs/2008ASPC..391..335F}

\bibitem[{{Gaia Collaboration} {et~al.}(2022){Gaia Collaboration}, Vallenari,
  Brown, Prusti, de~Bruijne, Arenou, Babusiaux, Biermann, Creevey, Ducourant,
  Evans, Eyer, Guerra, Hutton, Jordi, Klioner, Lammers, Lindegren, Luri,
  Mignard, Panem, Pourbaix, Randich, Sartoretti, Soubiran, Tanga, Walton,
  Bailer-Jones, Bastian, Drimmel, Jansen, Katz, Lattanzi, van Leeuwen, Bakker,
  Cacciari, Castañeda, De~Angeli, Fabricius, Fouesneau, Frémat, Galluccio,
  Guerrier, Heiter, Masana, Messineo, Mowlavi, Nicolas, Nienartowicz, Pailler,
  Panuzzo, Riclet, Roux, Seabroke, Sordoørcit, Thévenin, Gracia-Abril,
  Portell, Teyssier, Altmann, Andrae, Audard, Bellas-Velidis, Benson, Berthier,
  Blomme, Burgess, Busonero, Busso, Cánovas, Carry, Cellino, Cheek,
  Clementini, Damerdji, Davidson, de~Teodoro, Nuñez~Campos, Delchambre,
  Dell'Oro, Esquej, Fernández-Hernández, Fraile, Garabato, García-Lario,
  Gosset, Haigron, Halbwachs, Hambly, Harrison, Hernández, Hestroffer,
  Hodgkin, Holl, Janßen, Jevardat~de Fombelle, Jordan, Krone-Martins,
  Lanzafame, Löffler, Marchal, Marrese, Moitinho, Muinonen, Osborne, Pancino,
  Pauwels, Recio-Blanco, Reylé, Riello, Rimoldini, Roegiers, Rybizki, Sarro,
  Siopis, Smith, Sozzetti, Utrilla, van Leeuwen, Abbas, Ábrahám,
  Abreu~Aramburu, Aerts, Aguado, Ajaj, Aldea-Montero, Altavilla, Álvarez,
  Alves, Anders, Anderson, Anglada~Varela, Antoja, Baines, Baker,
  Balaguer-Núñez, Balbinot, Balog, Barache, Barbato, Barros, Barstow,
  Bartolomé, Bassilana, Bauchet, Becciani, Bellazzini, Berihuete, Bernet,
  Bertone, Bianchi, Binnenfeld, Blanco-Cuaresma, Blazere, Boch, Bombrun,
  Bossini, Bouquillon, Bragaglia, Bramante, Breedt, Bressan, Brouillet,
  Brugaletta, Bucciarelli, Burlacu, Butkevich, Buzzi, Caffau, Cancelliere,
  Cantat-Gaudin, Carballo, Carlucci, Carnerero, Carrasco, Casamiquela,
  Castellani, Castro-Ginard, Chaoul, Charlot, Chemin, Chiaramida, Chiavassa,
  Chornay, Comoretto, Contursi, Cooper, Cornez, Cowell, Crifo, Cropper, Crosta,
  Crowley, Dafonte, Dapergolas, David, David, de~Laverny, De~Luise, De~March,
  De~Ridder, de~Souza, de~Torres, del Peloso, del Pozo, Delbo, Delgado,
  Delisle, Demouchy, Dharmawardena, Di~Matteo, Diakite, Diener, Distefano,
  Dolding, Edvardsson, Enke, Fabre, Fabrizio, Faigler, Fedorets, Fernique,
  Fienga, Figueras, Fournier, Fouron, Fragkoudi, Gai, Garcia-Gutierrez,
  Garcia-Reinaldos, García-Torres, Garofalo, Gavel, Gavras, Gerlach, Geyer,
  Giacobbe, Gilmore, Girona, Giuffrida, Gomel, Gomez, González-Núñez,
  González-Santamaría, González-Vidal, Granvik, Guillout, Guiraud,
  Gutiérrez-Sánchez, Guy, Hatzidimitriou, Hauser, Haywood, Helmer, Helmi,
  Sarmiento, Hidalgo, Hilger, Hładczuk, Hobbs, Holland, Huckle, Jardine,
  Jasniewicz, Jean-Antoine~Piccolo, Jiménez-Arranz, Jorissen,
  Juaristi~Campillo, Julbe, Karbevska, Kervella, Khanna, Kontizas, Kordopatis,
  Korn, Kóspál, Kostrzewa-Rutkowska, Kruszyńska, Kun, Laizeau, Lambert,
  Lanza, Lasne, Le~Campion, Lebreton, Lebzelter, Leccia, Leclerc,
  Lecoeur-Taibi, Liao, Licata, Lindstrøm, Lister, Livanou, Lobel, Lorca, Loup,
  Madrero~Pardo, Magdaleno~Romeo, Managau, Mann, Manteiga, Marchant, Marconi,
  Marcos, Marcos~Santos, Marín~Pina, Marinoni, Marocco, Marshall, Polo,
  Martín-Fleitas, Marton, Mary, Masip, Massari, Mastrobuono-Battisti, Mazeh,
  McMillan, Messina, Michalik, Millar, Mints, Molina, Molinaro, Molnár,
  Monari, Monguió, Montegriffo, Montero, Mor, Mora, Morbidelli, Morel, Morris,
  Muraveva, Murphy, Musella, Nagy, Noval, Ocaña, Ogden, Ordenovic, Osinde,
  Pagani, Pagano, Palaversa, Palicio, Pallas-Quintela, Panahi, Payne-Wardenaar,
  Peñalosa~Esteller, Penttilä, Pichon, Piersimoni, Pineau, Plachy, Plum,
  Poggio, Prša, Pulone, Racero, Ragaini, Rainer, Raiteri, Rambaux, Ramos,
  Ramos-Lerate, Re~Fiorentin, Regibo, Richards, Rios~Diaz, Ripepi, Riva, Rix,
  Rixon, Robichon, Robin, Robin, Roelens, Rogues, Rohrbasser, Romero-Gómez,
  Rowell, Royer, Ruz~Mieres, Rybicki, Sadowski, Sáez~Núñez,
  Sagristà~Sellés, Sahlmann, Salguero, Samaras, Sanchez~Gimenez, Sanna,
  Santoveña, Sarasso, Schultheis, Sciacca, Segol, Segovia, Ségransan, Semeux,
  Shahaf, Siddiqui, Siebert, Siltala, Silvelo, Slezak, Slezak, Smart, Snaith,
  Solano, Solitro, Souami, Souchay, Spagna, Spina, Spoto, Steele,
  Steidelmüller, Stephenson, Süveges, Surdej, Szabados, Szegedi-Elek, Taris,
  Taylo, Teixeira, Tolomei, Tonello, Torra, Torra, Torralba~Elipe, Trabucchi,
  Tsounis, Turon, Ulla, Unger, Vaillant, van Dillen, van Reeven, Vanel,
  Vecchiato, Viala, Vicente, Voutsinas, Weiler, Wevers, Wyrzykowski, Yoldas,
  Yvard, Zhao, Zorec, Zucker, \&
  Zwitter}]{gaiacollaborationGaiaDataRelease2022}
{Gaia Collaboration}, Vallenari, A., Brown, A. G.~A., {et~al.} 2022, Gaia
  {Data} {Release} 3: {Summary} of the content and survey properties, Tech.
  rep., \dodoi{10.48550/arXiv.2208.00211}

\bibitem[{García-Hernández {et~al.}(2011{\natexlab{a}})García-Hernández,
  Kameswara~Rao, \& Lambert}]{garcia-hernandezAreC60Molecules2011}
García-Hernández, D.~A., Kameswara~Rao, N., \& Lambert, D.~L.
  2011{\natexlab{a}}, The Astrophysical Journal, 729, 126,
  \dodoi{10.1088/0004-637X/729/2/126}

\bibitem[{García-Hernández {et~al.}(2011{\natexlab{b}})García-Hernández,
  Rao, \& Lambert}]{garcia-hernandezDustCoronaeBorealis2011}
García-Hernández, D.~A., Rao, N.~K., \& Lambert, D.~L. 2011{\natexlab{b}},
  The Astrophysical Journal, 739, 37, \dodoi{10.1088/0004-637X/739/1/37}

\bibitem[{García-Hernández {et~al.}(2013)García-Hernández, Rao, \&
  Lambert}]{garcia-hernandezDustCoronaeBorealis2013}
---. 2013, The Astrophysical Journal, 773, 107,
  \dodoi{10.1088/0004-637X/773/2/107}

\bibitem[{Harper {et~al.}(2018)Harper, Runyan, Dowell, Wirth, Amato, Ames,
  Amiri, Banks, Bartels, Benford, Berthoud, Buchanan, Casey, Chapman, Chuss,
  Cook, Derro, Dotson, Evans, Fixsen, Gatley, Guerra, Halpern, Hamilton,
  Hamlin, Hansen, Heimsath, Hermida, Hilton, Hirsch, Hollister, Hostetter,
  Irwin, Jhabvala, Jhabvala, Kastner, Kovács, Lin, Loewenstein, Looney,
  Lopez-Rodriguez, Maher, Michail, Miller, Moseley, Novak, Pernic, Rennick,
  Rhody, Sandberg, Sandford, Santos, Shafer, Sharp, Shirron, Siah, Silverberg,
  Sparr, Spotz, Staguhn, Toorian, Towey, Tuttle, Vaillancourt, Voellmer,
  Volpert, Wang, \& Wollack}]{harperHAWCFarInfraredCamera2018}
Harper, D.~A., Runyan, M.~C., Dowell, C.~D., {et~al.} 2018, Journal of
  Astronomical Instrumentation, 7, 1840008, \dodoi{10.1142/S2251171718400081}

\bibitem[{Hecht {et~al.}(1984)Hecht, Holm, Donn, \&
  Wu}]{hechtDustCoronaeBorealis1984}
Hecht, J.~H., Holm, A.~V., Donn, B., \& Wu, C.~C. 1984, The Astrophysical
  Journal, 280, 228, \dodoi{10.1086/161989}

\bibitem[{Henden {et~al.}(2015)Henden, Levine, Terrell, \&
  Welch}]{hendenAPASSLatestData2015}
Henden, A.~A., Levine, S., Terrell, D., \& Welch, D.~L. 2015, 225, 336.16.
\newblock \url{https://ui.adsabs.harvard.edu/abs/2015AAS...22533616H}

\bibitem[{Herter {et~al.}(2013)Herter, Vacca, Adams, Keller, Schoenwald,
  Hirsch, Wang, De~Buizer, Helton, \& Llorens}]{herterDataReductionEarly2013}
Herter, T.~L., Vacca, W.~D., Adams, J.~D., {et~al.} 2013, Publications of the
  Astronomical Society of the Pacific, 125, 1393, \dodoi{10.1086/674144}

\bibitem[{Iben {et~al.}(1996)Iben, Tutukov, \&
  Yungelson}]{ibenOriginCrBOther1996}
Iben, I., Tutukov, A.~V., \& Yungelson, L.~R. 1996, 96, 409.
\newblock \url{https://ui.adsabs.harvard.edu/abs/1996ASPC...96..409I}

\bibitem[{Kameswara~Rao {et~al.}(2004)Kameswara~Rao, Reddy, \&
  Lambert}]{kameswararaoRCoronaeBorealisStars2004}
Kameswara~Rao, N., Reddy, B.~E., \& Lambert, D.~L. 2004, Monthly Notices of the
  Royal Astronomical Society, 355, 855,
  \dodoi{10.1111/j.1365-2966.2004.08363.x}

\bibitem[{Karambelkar {et~al.}(2021)Karambelkar, Kasliwal, Tisserand, De,
  Anand, Ashley, Delacroix, Hankins, Jencson, Lau, McKenna, Moore, Ofek, Smith,
  Soria, Soon, Tinyanont, Travouillon, \&
  Yao}]{karambelkarCensusCoronaeBorealis2021}
Karambelkar, V.~R., Kasliwal, M.~M., Tisserand, P., {et~al.} 2021, The
  Astrophysical Journal, 910, 132, \dodoi{10.3847/1538-4357/abe5aa}

\bibitem[{Kraemer {et~al.}(2005)Kraemer, Sloan, Wood, Price, \&
  Egan}]{kraemerCrBCandidatesSmall2005}
Kraemer, K.~E., Sloan, G.~C., Wood, P.~R., Price, S.~D., \& Egan, M.~P. 2005,
  The Astrophysical Journal, 631, L147, \dodoi{10.1086/497427}

\bibitem[{Lambert {et~al.}(1990)Lambert, Rao, \&
  Giridhar}]{lambertHighResolutionSpectroscopy1990}
Lambert, D.~L., Rao, N.~K., \& Giridhar, S. 1990, Journal of Astrophysics and
  Astronomy, 11, 475, \dodoi{10.1007/BF02709762}

\bibitem[{Lambert {et~al.}(2001)Lambert, Rao, Pandey, \&
  Ivans}]{lambertInfraredSpaceObservatory2001}
Lambert, D.~L., Rao, N.~K., Pandey, G., \& Ivans, I.~I. 2001, The Astrophysical
  Journal, 555, 925, \dodoi{10.1086/321504}

\bibitem[{Lebouteiller {et~al.}(2011)Lebouteiller, Barry, Spoon, Bernard-Salas,
  Sloan, Houck, \& Weedman}]{lebouteillerCASSISCornellAtlas2011}
Lebouteiller, V., Barry, D.~J., Spoon, H. W.~W., {et~al.} 2011, The
  Astrophysical Journal Supplement Series, 196, 8,
  \dodoi{10.1088/0067-0049/196/1/8}

\bibitem[{Mainzer {et~al.}(2014)Mainzer, Bauer, Cutri, Grav, Masiero, Beck,
  Clarkson, Conrow, Dailey, Eisenhardt, Fabinsky, Fajardo-Acosta, Fowler,
  Gelino, Grillmair, Heinrichsen, Kendall, Kirkpatrick, Liu, Masci, McCallon,
  Nugent, Papin, Rice, Royer, Ryan, Sevilla, Sonnett, Stevenson, Thompson,
  Wheelock, Wiemer, Wittman, Wright, \&
  Yan}]{mainzerInitialPerformanceNEOWISE2014}
Mainzer, A., Bauer, J., Cutri, R.~M., {et~al.} 2014, The Astrophysical Journal,
  792, 30, \dodoi{10.1088/0004-637X/792/1/30}

\bibitem[{Renzini(1990)}]{renziniEvolutionaryScenariosCrB1990}
Renzini, A. 1990, 11, 549.
\newblock \url{https://ui.adsabs.harvard.edu/abs/1990ASPC...11..549R}

\bibitem[{Riello {et~al.}(2021)Riello, De~Angeli, Evans, Montegriffo, Carrasco,
  Busso, Palaversa, Burgess, Diener, Davidson, Rowell, Fabricius, Jordi,
  Bellazzini, Pancino, Harrison, Cacciari, van Leeuwen, Hambly, Hodgkin,
  Osborne, Altavilla, Barstow, Brown, Castellani, Cowell, De~Luise, Gilmore,
  Giuffrida, Hidalgo, Holland, Marinoni, Pagani, Piersimoni, Pulone, Ragaini,
  Rainer, Richards, Sanna, Walton, Weiler, \& Yoldas}]{rielloGaiaEarlyData2021}
Riello, M., De~Angeli, F., Evans, D.~W., {et~al.} 2021, Astronomy and
  Astrophysics, 649, A3, \dodoi{10.1051/0004-6361/202039587}

\bibitem[{Shappee {et~al.}(2014)Shappee, Prieto, Stanek, Kochanek, Holoien,
  Jencson, Basu, Beacom, Szczygiel, Pojmanski, Brimacombe, Dubberley, Elphick,
  Foale, Hawkins, Mullins, Rosing, Ross, \&
  Walker}]{shappeeAllSkyAutomated2014}
Shappee, B., Prieto, J., Stanek, K.~Z., {et~al.} 2014, 223, 236.03.
\newblock \url{https://ui.adsabs.harvard.edu/abs/2014AAS...22323603S}

\bibitem[{Su {et~al.}(2017)Su, De~Buizer, Rieke, Krivov, Löhne, Marengo,
  Stapelfeldt, Ballering, \& Vacca}]{suInner25Au2017}
Su, K. Y.~L., De~Buizer, J.~M., Rieke, G.~H., {et~al.} 2017, The Astronomical
  Journal, 153, 226, \dodoi{10.3847/1538-3881/aa696b}

\bibitem[{Tisserand(2012)}]{tisserandTrackingCoronaeBorealis2012}
Tisserand, P. 2012, Astronomy \&amp; Astrophysics, Volume 539, id.A51,
  {\textless}NUMPAGES{\textgreater}16{\textless}/NUMPAGES{\textgreater} pp.,
  539, A51, \dodoi{10.1051/0004-6361/201117874}

\bibitem[{Tisserand {et~al.}(2020)Tisserand, Clayton, Bessell, Welch, Kamath,
  Wood, Wils, Wyrzykowski, Mróz, \& Udalski}]{tisserandPlethoraNewCoronae2020}
Tisserand, P., Clayton, G.~C., Bessell, M.~S., {et~al.} 2020, Astronomy and
  Astrophysics, 635, A14, \dodoi{10.1051/0004-6361/201834410}

\bibitem[{Tonry {et~al.}(2018)Tonry, Denneau, Heinze, Stalder, Smith, Smartt,
  Stubbs, Weiland, \& Rest}]{tonryATLASHighcadenceAllsky2018}
Tonry, J.~L., Denneau, L., Heinze, A.~N., {et~al.} 2018, Publications of the
  Astronomical Society of the Pacific, 130, 064505,
  \dodoi{10.1088/1538-3873/aabadf}

\bibitem[{Wilson {et~al.}(2003)Wilson, Eikenberry, Henderson, Hayward, Carson,
  Pirger, Barry, Brandl, Houck, Fitzgerald, \&
  Stolberg}]{wilsonWideFieldInfraredCamera2003a}
Wilson, J.~C., Eikenberry, S.~S., Henderson, C.~P., {et~al.} 2003, 4841, 451,
  \dodoi{10.1117/12.460336}

\bibitem[{Wright(1989)}]{wrightFractalDustGrains1989}
Wright, E.~L. 1989, The Astrophysical Journal, 346, L89, \dodoi{10.1086/185586}

\bibitem[{Wright {et~al.}(2010)Wright, Eisenhardt, Mainzer, Ressler, Cutri,
  Jarrett, Kirkpatrick, Padgett, McMillan, Skrutskie, Stanford, Cohen, Walker,
  Mather, Leisawitz, Gautier, McLean, Benford, Lonsdale, Blain, Mendez, Irace,
  Duval, Liu, Royer, Heinrichsen, Howard, Shannon, Kendall, Walsh, Larsen,
  Cardon, Schick, Schwalm, Abid, Fabinsky, Naes, \&
  Tsai}]{wrightWidefieldInfraredSurvey2010}
Wright, E.~L., Eisenhardt, P. R.~M., Mainzer, A.~K., {et~al.} 2010, The
  Astronomical Journal, 140, 1868, \dodoi{10.1088/0004-6256/140/6/1868}

\end{thebibliography}

\clearpage

\appendix{\label{appendix}}
\setcounter{table}{0}
\renewcommand{\thetable}{A\arabic{table}}
\setcounter{figure}{0}
\renewcommand{\thefigure}{A\arabic{figure}}
%\appendixpage

%\section{Supporting Information Tables}

Here we report magnitudes for the reference source TYC 4380 226 1
(Table \ref{table:ref_star})
and color-corrected mid-infrared flux values for {\it IRAS},
{\it AKARI}, and {\it WISE} data of IRAS\,00450+7401 (Table \ref{table:color_correc}).

\begin{table}[!h]
\caption{{\large TYC 4380 226 1 Magnitudes}
\label{table:ref_star}}
\begin{center}
\begin{tabular}{ccc}
\hline 
\hline
Instrument/Mission & Bandpass & Magnitude (Vega) \\
\hline
APASS DR9 & B & 12.777$\pm$0.071\\
APASS DR9 & V & 12.486$\pm$0.052\\
ASAS-SN & V & 12.43$\pm$0.022 \\
{\it Gaia} DR3 & V & 12.527$\pm$0.03\\
{\it Gaia} DR3 & R & 12.332$\pm$0.03\\
{\it Gaia} DR3 & I & 12.120$\pm$0.04\\
\hline
\end{tabular}
\\
{\it Note} $-$ Data used: AAVSO Photometric All Sky Survey (APASS; \citealt{hendenAPASSLatestData2015}), All-Sky Automated Survey for Supernovae (ASAS-SN; \citealt{shappeeAllSkyAutomated2014}), and {\it Gaia} DR3 (\citealt{collaborationGaiaMission2016}; \citealt{gaiacollaborationGaiaDataRelease2022}). 
{\it Gaia} G, BP, and RP magnitudes are converted to V, R, and I magnitudes
following \citet{rielloGaiaEarlyData2021}.
An average of the three \textit{V-}band magnitudes was adopted.
%{\it Note} 
\end{center}
\end{table}
%%%%%%%%%%%%%%%%%%%%%%%%%%%%%%%%%%%%%%%%%%%%%%%%%%%%%%%%%%%%%%%%%%%%%%%%%%%%%%%%%%%

\begin{table}[!h]
\caption{{\large Color-corrected IRAS\,00450+7401 Infrared Flux Values}
\label{table:color_correc}}
\begin{center}
\begin{tabular}{cccc}
\hline 
\hline
Mission & Bandpass & Flux (Jy) & Epoch\\
\hline

{\it IRAS} & 12\,$\mu$m & 3.36 $\pm$ 0.13 & 1983\\
{\it IRAS} & 25\,$\mu$m & 1.59 $\pm$ 0.08 & 1983\\
{\it IRAS} & 60\,$\mu$m & 0.37 $\pm$ 0.05 & 1983\\
{\it AKARI} & 9\,$\mu$m & 2.36 $\pm$ 0.02 & 2006\\
{\it AKARI} & 18\,$\mu$m & 1.39 $\pm$ 0.01 & 2006\\
{\it WISE} & 11.56\,$\mu$m & 1.97 $\pm$ 0.02 & 2010\\
{\it WISE} & 22.09\,$\mu$m & 1.10 $\pm$ 0.02 & 2010\\

\hline
\end{tabular}
%{\it Note} $-$ 
\end{center}
\end{table}
%%%%%%%%%%%%%%%%%%%%%%%%%%%%%%%%%%%%%%%%%%%%%%%%%%%%%%%%%%%%%%%%%%%%%%%%%%%%%%%%%%%

\end{document}